\documentclass[12pt, draftcls, onecolumn]{IEEEtran}
\usepackage[T1]{fontenc}
\usepackage[utf8]{inputenc}
\usepackage{mathtools}
\usepackage{amsthm, amssymb}
\usepackage[binary-units=true]{siunitx}
\DeclareSIUnit{\bits}{bits}
\usepackage{subcaption} 
\usepackage[noadjust]{cite}
\usepackage{tikz}
\usepackage[draft]{hyperref}
\usepackage{pgfplots}
\usetikzlibrary{pgfplots.groupplots}
\pgfplotsset{compat=1.17}
\usepackage[acronym]{glossaries}
\usepackage[ruled,vlined]{algorithm2e}
\usepackage{hhline}

\newacronym{3GPP}{3GPP}{3rd Generation Partnership Project}
\newacronym{ACM}{ACM}{adaptive coding and modulation}
\newacronym{ACLR}{ACLR}{adjacent channel leakage ratio}
\newacronym{ADC}{ADC}{analog-to-digital conversion}
\newacronym{AGC}{AGC}{automatic gain control}
\newacronym{AWGN}{AWGN}{additive white Gaussian noise}
\newacronym{BER}{BER}{bit error rate}
\newacronym{BS}{BS}{base station}
\newacronym{BLER}{BLER}{block error rate}
\newacronym{BCE}{BCE}{binary cross-entropy}
\newacronym{BICM}{BICM}{bit-interleaved coded modulation}
\newacronym{BMD}{BMD}{bit-metric decoding}
\newacronym{BP}{BP}{backpropagation}
\newacronym{BPTT}{BPTT}{backpropagation through time}
\newacronym{CE}{CE}{cross-entropy}
\newacronym{CFO}{CFO}{carrier frequency offset}
\newacronym{CNN}{CNN}{convolutional neural network}
\newacronym[longplural={cyclic prefixes}]{CP}{CP}{cyclic prefix}
\newacronym{CCDF}{CCDF}{complementary cumulative distribution function}
\newacronym{CSI}{CSI}{channel state information}
\newacronym{DAC}{DAC}{digital-to-analog conversion}
\newacronym{DPD}{DPD}{digital pre-distortion}
\newacronym{DFT}{DFT}{discrete Fourier transform}
\newacronym{DL}{DL}{deep learning}
\newacronym{ELU}{ELU}{exponential linear unit}
\newacronym{FFT}{FFT}{fast Fourier transform}
\newacronym{FBS}{FBS}{frequency baseband symbol}
\newacronym{GAN}{GAN}{generative adversarial network}
\newacronym{GRU}{GRU}{gated recurrent unit}
\newacronym{iid}{i.i.d.\@}{independent and identically distributed}
\newacronym{IFFT}{IFFT}{inverse fast Fourier transform}
\newacronym{IDFT}{IDFT}{inverse discrete Fourier transform}
\newacronym{KL}{KL}{Kullback-Leibler}
\newacronym{LLR}{LLR}{log-likelihood ratio}
\newacronym{LSTM}{LSTM}{long short-term memory}
\newacronym{LDPC}{LDPC}{low-density parity-check}
\newacronym{LMMSE}{LMMSE}{linear minimum mean squared error}
\newacronym{MDP}{MDP}{Markov decision process}
\newacronym{ML}{ML}{machine learning}
\newacronym{MLP}{MLP}{multilayer perceptron}
\newacronym{MIMO}{MIMO}{multiple-input multiple-output}
\newacronym{MU-MIMO}{MU-MIMO}{multi-user multiple-input multiple-output}
\newacronym{MU}{MU}{multi-user}
\newacronym{MSE}{MSE}{mean squared error}
\newacronym{NN}{NN}{neural network}
\newacronym{NR}{NR}{new radio}
\newacronym{NLOS}{NLOS}{non-line of sight}
\newacronym{OFDM}{OFDM}{orthogonal frequency-division multiplexing}
\newacronym{pdf}{pdf}{probability density function}
\newacronym{pmf}{pmf}{probability mass function}
\newacronym{PA}{PA}{power amplifier}
\newacronym{PAPR}{PAPR}{peak-to-average power ratio}
\newacronym[longplural={power spectral densities}]{PSD}{PSD}{power spectral density}
\newacronym{PRT}{PRT}{peak reduction tone}
\newacronym{QPSK}{QPSK}{quadrature phase-shift keying}
\newacronym{QAM}{QAM}{quadrature amplitude modulation}
\newacronym{PSNR}{PSNR}{Peak Signal to Noise Ratio}
\newacronym{RBF}{RBF}{Rayleigh block-fading}
\newacronym{RB}{RB}{resource block}
\newacronym{RE}{RE}{resource element}
\newacronym{RG}{RG}{resource grid\newacronym{RE}{RE}{resource element}}
\newacronym{ReLU}{ReLU}{rectified linear unit}
\newacronym{RTN}{RTN}{radio transformer network}
\newacronym{RL}{RL}{reinforcement learning}
\newacronym{RNN}{RNN}{recurrent neural network}
\newacronym{SFO}{SFO}{sampling frequency offset}
\newacronym{SER}{SER}{symbol error rate}
\newacronym{SNR}{SNR}{signal-to-noise ratio}
\newacronym{SINR}{SINR}{signal-to-interference-plus-noise ratio}
\newacronym{SGD}{SGD}{stochastic gradient descent}
\newacronym{SISO}{SISO}{single-input single-output}
\newacronym{SIMO}{SIMO}{single-input multiple-output}
\newacronym{SU}{SU}{single-user}
\newacronym{TDD}{TDD}{time-division duplexing}
\newacronym{TR}{TR}{tone reservation}
\newacronym{UE}{UE}{user equipment}
\newacronym{UMi}{UMi}{urban microcell}
\newacronym{wrt}{w.r.t.\@}{with respect to}

\renewcommand{\vec}[1]{\mathbf{#1}}
\newcommand{\vecs}[1]{\boldsymbol{#1}}

\newcommand{\bv}{\vec{b}}
\newcommand{\cv}{\vec{c}}

\newcommand{\hv}{\vec{h}}

\newcommand{\mv}{\vec{m}}

\newcommand{\pv}{\vec{p}}

\newcommand{\uv}{\vec{u}}

\newcommand{\xv}{\vec{x}}
\newcommand{\yv}{\vec{y}}
\newcommand{\zv}{\vec{z}}

\newcommand{\thetav}{\vecs{\theta}}
\newcommand{\psiv}{\vecs{\psi}}

\newcommand{\Bm}{\vec{B}}

\newcommand{\Fm}{\vec{F}}

\newcommand{\Hm}{\vec{H}}
\newcommand{\Id}{\vec{I}}

\newcommand{\Mm}{\vec{M}}
\newcommand{\Nm}{\vec{N}}

\newcommand{\Vm}{\vec{V}}
\newcommand{\Wm}{\vec{W}}
\newcommand{\Xm}{\vec{X}}
\newcommand{\Ym}{\vec{Y}}

\newcommand{\Sigmam}{\vecs{\Sigma}}

\newcommand{\Cc}{{\cal C}}

\newcommand{\Nc}{{\cal N}}

\newcommand{\CC}{\mathbb{C}}

\newcommand{\RR}{\mathbb{R}}

\newcommand{\htp}{^{\mathsf{H}}}
\newcommand{\tp}{^{\mathsf{T}}}

\newcommand{\LB}{\left(}
\newcommand{\RB}{\right)}
\newcommand{\LP}{\left\{}
\newcommand{\RP}{\right\}}
\newcommand{\LSB}{\left[}
\newcommand{\RSB}{\right]}

\newcommand{\EE}{{\mathbb{E}}}

\newlength{\dhatheight}

\newcommand{\gpeak}{\gamma_{\text{peak}}}
\newcommand{\bleak}{\beta_{\text{leak}}}

\begin{document}
\title{Learning OFDM Waveforms with PAPR and ACLR Constraints}

\author{Mathieu~Goutay,~
\IEEEmembership{Student Member,~IEEE},
        Fayçal~Ait~Aoudia,
       	~\IEEEmembership{Member,~IEEE},
        ~Jakob~Hoydis,
        ~\IEEEmembership{Senior~Member,~IEEE},
        and Jean-Marie~Gorce,
        ~\IEEEmembership{Senior~Member,~IEEE}
\thanks{Mathieu Goutay is with Nokia Bell Labs, Paris-Saclay, 91620 Nozay,
France, and also with the CITI Lab, INSA Lyon, Inria, Université de Lyon,
69100 Villeurbanne, France (e-mail: mathieu.goutay@nokia.com).}
\thanks{Fayçal Ait Aoudia and Jakob Hoydis were with Nokia Bell Labs, Paris-Saclay, 91620 Nozay,
France. They are now with NVIDIA, 06906 Sophia Antipolis, France (e-mail:
\{faitaoudia, jhoydis\}@nvidia.com).}
\thanks{Jean-Marie Gorce is with the CITI Lab, INSA Lyon, Inria, Université de
Lyon, 69100 Villeurbanne, France (e-mail: jean-marie.gorce@insa-lyon.fr).}}

\maketitle

\begin{abstract}

An attractive research direction for future communication systems is the design of new waveforms that can both support high throughputs and present advantageous signal characteristics.
Although most modern systems use \gls{OFDM} for its efficient equalization, this waveform suffers from multiple limitations such as a high \gls{ACLR} and high \gls{PAPR}.
In this paper, we propose a learning-based method to design \gls{OFDM}-based waveforms that satisfy selected constraints while maximizing an achievable information rate. 
To that aim, we model the transmitter and the receiver as \glspl{CNN} that respectively implement a high-dimensional modulation scheme and perform the detection of the transmitted bits.
This leads to an optimization problem that is solved using the augmented Lagrangian method.
Evaluation results show that the end-to-end system is able to satisfy target \gls{PAPR} and \gls{ACLR} constraints and allows significant throughput gains compared to a \gls{TR} baseline.
An additional advantage is that no dedicated pilots are needed.


\end{abstract}

\glsresetall

\section{Introduction} 
\label{sec:introduction}

The next generation of cellular networks will need to support a growing number of different services and devices~\cite{9040431}. 
Along with higher throughputs, future communication systems are expected to satisfy new requirements on the signal characteristics.
For example, a higher number of connected devices suggests that the available spectrum should be more efficiently shared among users, challenging the need for guard bands.
Moreover, the \gls{PA} nonlinearities in sub-\si{THz} systems lead to the use of large power backoffs to prevent distortions caused by high-amplitude signals.
A key research direction to address these issues is the design of new waveforms.
Among possible candidates, \gls{OFDM} is already used in most modern communication systems, including 4G, 5G, and Wi-Fi.
The main benefits of \gls{OFDM} are a very efficient hardware implementation and a single-tap equalization at the receiver.
However, conventional \gls{OFDM} suffers from multiple drawbacks, including the addition of pilots for channel equalization, a high sensitivity to Doppler spread, and both a high \gls{PAPR} and \gls{ACLR}, which might hinder its use in beyond-5G systems.

Future base stations and user equipments are expected to be equipped with dedicated \gls{ML} accelerators~\cite{hoydis20216g} that enable efficient implementation of \gls{NN}-based components.
In this paper, we take advantage of such capabilities and propose a new learning-based approach to \gls{OFDM}-based waveform design.
This approach is based on a \gls{CNN} transmitter that implements a high-dimension modulation scheme and a CNN-based receiver that computes \glspl{LLR} on the transmitted bits.
Both transceivers operate on top of OFDM to benefit from its efficient hardware implementation and process OFDM symbols instead of individual \glspl{RE}.
We derive a training procedure is derived which allows to both maximize an achievable information rate and to offset OFDM drawbacks by defining specific optimization constraints.
The end-to-end system training is performed through \gls{SGD}, and therefore the achievable rate and the constraints need to be expressed as functions that can be evaluated and differentiated during training.

In the following, we focus on designing OFDM-based waveforms that enable pilotless transmissions and satisfy \gls{PAPR} and \gls{ACLR} constraints.
The end-to-end system is benchmarked against a close to ideal implementation of a \gls{TR} baseline, in which a number of subcarriers are used to generate peak-reduction signals.
Both systems are evaluated on a \gls{3GPP}-compliant channel model, and the baseline uses a pilot configuration supported by the 5G \gls{NR} specifications.
The end-to-end system, on the contrary, does not use any pilots and learns a high-dimensional modulation that enables accurate detection at the receiver.
Evaluation results show that the learning-based system allows to meet \gls{PAPR} and \gls{ACLR} targets and enables throughput gains ranging from 3\% to 30\% compared to a baseline with similar characteristics.
To get insight into how the proposed system reduces the \gls{PAPR} and \gls{ACLR} while maintaining high rates, we have carried out a detailed study of the learned high-dimensional modulation scheme.
We have found out that \gls{ACLR} and \gls{PAPR} reduction is achieved through a combination of spectral filtering, uneven energy allocation across subcarriers, and positional adjustments of constellation symbols.
To the best of our knowledge, this method is the first \gls{ML}-based approach that jointly maximizes an information rate of OFDM transmissions and allows to set \gls{PAPR} and \gls{ACLR} targets.\footnote{Part of this work has been published in~\cite{goutay2021endtoend}.}

\subsection*{Related literature}

To counteract the drawbacks of OFDM-based waveforms, previous works suggested filtering the analog signal, to modify the constellation used for modulation, or to inject custom signals on reserved subcarriers.
The first approach is known as iterative clipping and filtering, and iterates between clipping in the time-domain  and filtering in the frequency domain to constrain the signal amplitude and possibly the spectral leakage~\cite{wang2010optimized}.
The second scheme, named active constellation extension, extends the outer symbols of a constellation to reduce the signal \gls{PAPR} at the cost of an increased power consumption~\cite{krongold2003reduction}.
Finally, the third technique reserves a subset of available subcarriers to create a peak-cancelling signal, and is referred to as tone reservation (TR)~\cite{1261335}.

Motivated by the success of \gls{ML} in other physical layer tasks, such as channel coding~\cite{7852251} and wireless communications~\cite{8839651}, multiple works suggested replacing existing algorithms with \gls{ML} components. 
For example, an \gls{NN} was used to generate the constellation extension vectors in~\cite{8928056}.
Similarly, a \gls{TR} algorithm was unfolded as \gls{NN} layers in~\cite{wang2020novel, wang2021model, 8928103}.
It has also been proposed to model the communication system as an autoencoder that is trained to both minimize the \gls{PAPR} and symbol error rate~\cite{8240644}.
Although closer to our approach, the proposed scheme operates on symbols, meaning that the bit mapping and demapping have to be implemented separately, and has only been evaluated on a simple Rayleigh fading channel.
Moreover, the time-domain signal is not oversampled, which is required to obtain an accurate representation of the underlying waveform for \gls{PAPR} calculations~\cite{922754}.
Finally, none of these works allow setting precise \gls{PAPR} and \gls{ACLR} targets, which means that the tradeoff between the \gls{PAPR}, \gls{ACLR}, and spectral efficiency can not be accurately controlled.

Regarding the end-to-end training of communication systems, numerous works proposed to leverage this technique to design transceivers aimed at different channels.
For example, the design of constellation geometries to achieve pilotless and \gls{CP}-less communication over OFDM channels was studied in~\cite{pilotless20}, while the design of transmit and receive filters for single-carrier systems under spectral and \gls{PAPR} constraints was presented in~\cite{aoudia2021endtoend}.
Other works in similar directions include the design of a \gls{DL}-aided multicarrier system for fading channels~\cite{9271932} and the design of \gls{NN}-based transceivers for optical fiber communications~\cite{8433895}.

\bigskip

\textbf{Notations :} 
$\RR$ ($\CC$) denotes the set of real (complex) numbers.
Tensors and matrices are denoted by bold upper-case letters and vectors are denoted by bold lower-case letters.
We respectively denote by $\mv_a$ and $m_{a,b}$ the vector and scalar formed by slicing the matrix $\Mm$ along its first and second dimension.
$(\cdot)\tp$, $(\cdot)\htp$, and $(\cdot)^*$ denote the transpose, conjugate transpose, and element-wise conjugate operator, respectively.
We denote by $\odot$ and $\oslash$ the element-wise multiplication and division operator, also known as Hadamard product and division.
$I(\xv; \yv)$ and $P(\xv, \yv)$ respectively represent the mutual information and joint conditional probability of $\xv$ and $\yv$.
Finally, the imaginary unit is denoted by $j$, such that $j^2 = -1$.

\section{Problem Statement} 
\label{sec:problem_positioning}

A \gls{SISO} system using OFDM is considered. 
In this section, the OFDM channel model is first presented, and the expressions of the signal waveform and spectrum are derived.
The \gls{ACLR} and \gls{PAPR} metrics typically used to characterize the analogue signal are then detailed. 
Finally, a close to ideal implementation of a \gls{TR} baseline is introduced, where a subset of subcarriers are reserved to minimize the signal \gls{PAPR} and pilots are transmitted to estimate the channel.

\subsection{System model}

\subsubsection{Channel model}
 
\begin{figure}[ht]
    \centering
    \begin{minipage}[b]{0.45\linewidth}
        \centering
    \includegraphics[height=100pt]{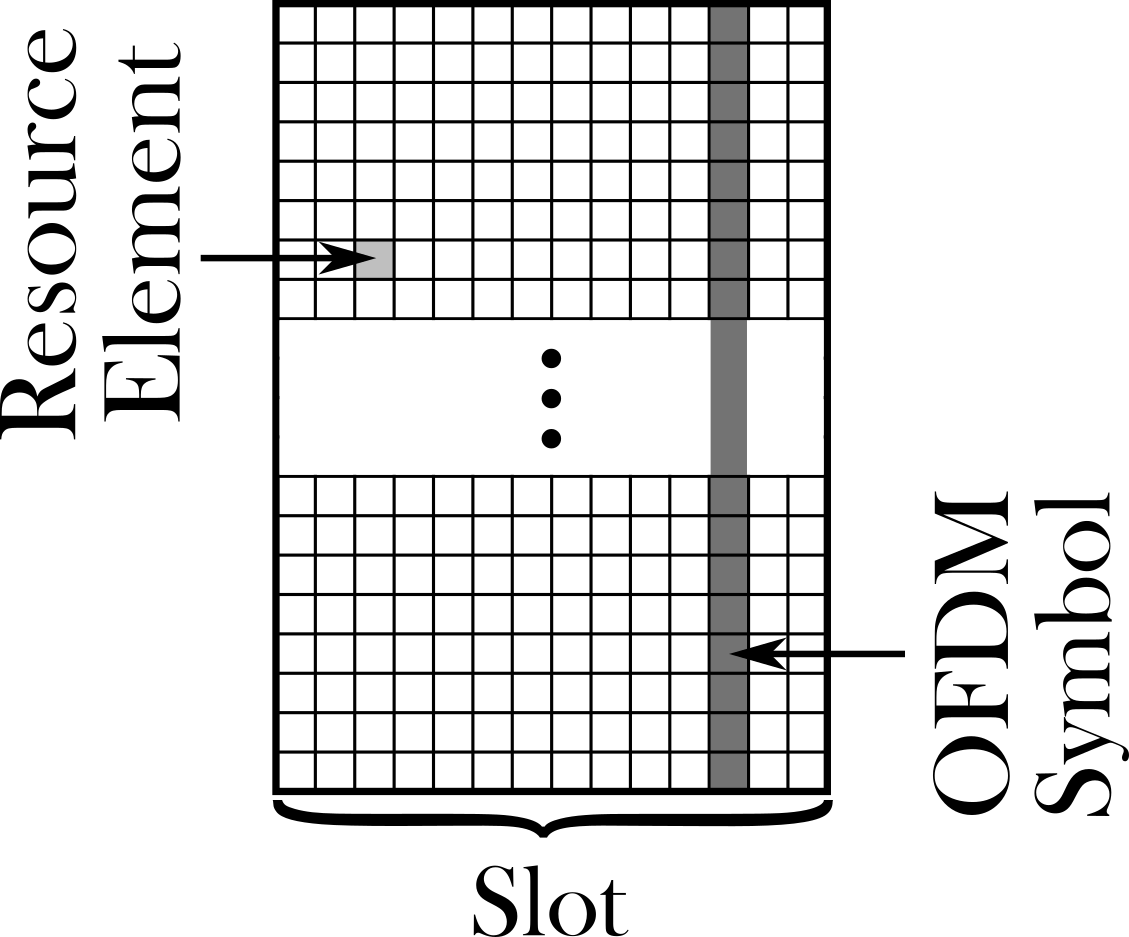}
    \caption{The frequency-time resource grid.}
    \label{fig:RG}
    \end{minipage}
    \hspace{0pt}
    \begin{minipage}[b]{0.45\linewidth}
    \centering
    \includegraphics[height=100pt]{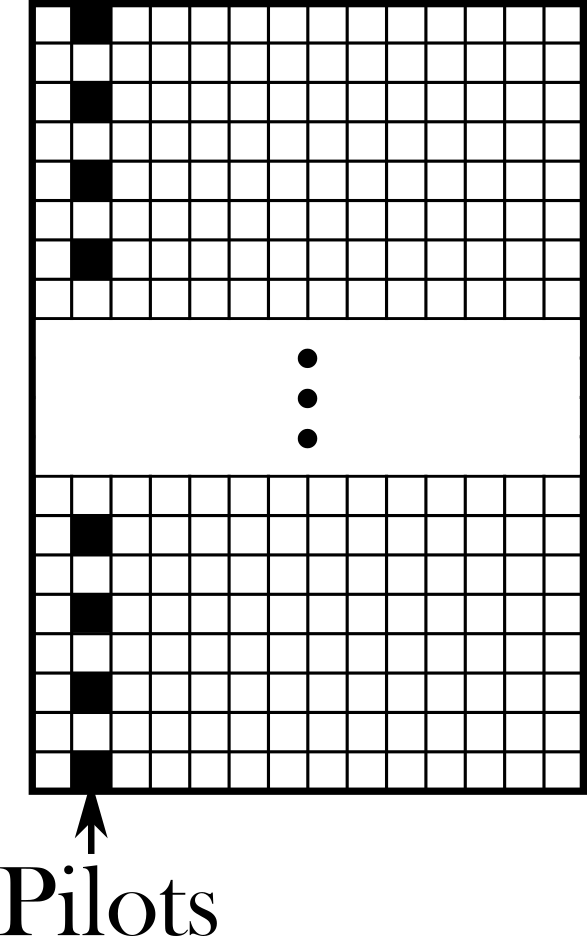}
    \caption{Pilot pattern used by the baseline.} 
    \label{fig:pilots}
    \end{minipage}
    \end{figure}

\sloppy Signals are transmitted over $N$ subcarriers and one time slot, which consists of $M=14$ adjacent OFDM symbols.
Each time-frequency position is called an \gls{RE} and the entire grid is referred to as the \gls{RG}, as depicted in Fig~\ref{fig:RG}. 
The subcarriers are indexed by the set $\mathcal{N}= \LP -\frac{N -1}{2}, \cdots, \frac{N -1}{2} \RP $, with $N$ assumed odd for convenience.
When using a \gls{CP}, the OFDM channel can be expressed as
\begin{align}
    \Ym = \Hm \odot \Xm + \Nm
    \label{eq:OFDM_channel}
\end{align}
where $\Xm \in \mathbb{C}^{M \times N}$ and $\Ym \in \mathbb{C}^{M \times N}$ respectively represent the matrix of sent and received \glspl{FBS}, $\Hm \in \mathbb{C}^{M \times N}$ is the matrix of channel coefficients, and {$\Nm~\in~\mathbb{C}^{M \times N}$} is the additive Gaussian noise matrix such that each element has a variance $\sigma^2$.
We consider a slow-varying environment so that the channel can be assumed constant over the duration of a slot.
The matrix of bits to be transmitted on the OFDM symbol $m$ is denoted $\Bm_{m} = \left[  \bv_{m,1}, \cdots, \bv_{m, N} \right]\tp$, where $\bv_{m, n}\in \{0,1\}^{K}, 1 \leq m \leq M, 1 \leq n \leq N$,  is a vector of bits to be sent and $K$ is the number of bits per channel use. 
The transmitter modulates each $\Bm_{m}$ onto the \glspl{FBS} $\xv_{m} \in \CC^{N}$, which are mapped on the orthogonal subcarriers to form the spectrum
\begin{align}
    S_{m}(f) = \sum_{n \in \mathcal{N}} x_{m, n} \frac{1}{\sqrt{\Delta_f}}\text{sinc} \left( \frac{f}{\Delta_f} - n \right)
\end{align}
where $\Delta_f$ is the subcarrier spacing.
The corresponding time-domain signal is 
\begin{align}
    s_{m}(t) = \sum_{n\in \mathcal{N}} x_{m, n} \frac{1}{\sqrt{T}} \text{rect} \left( \frac{t}{T}-m \right) e^{i 2 \pi n \frac{t-mT}{T}  }
\end{align}
where $ T = \frac{1}{\Delta_f}$ is the duration of an OFDM symbol without \gls{CP}.
To avoid intersymbol interference in multipath fading channels, \glspl{CP} of durations at least equal to the longest path need to be inserted.
If we denote by $T^{\text{CP}}$ the length of this CP, the transmitted signal is expressed as
\begin{align}
s_m^{\text{CP}}(t) = \left\{
    \begin{array}{ll}
        \frac{\sqrt{T}}{\sqrt{T+T^{\text{CP}}}} s_m(t-mT^{\text{CP}}) & \mbox{if \quad} t \text{ mod } T+T^{\text{CP}} \geq T_{\text{CP}}\\
        \frac{\sqrt{T}}{\sqrt{T+T^{\text{CP}}}} s_m(t-mT^{\text{CP}} +T) & \mbox{if \quad} t \text{ mod } T+T^{\text{CP}} < T_{\text{CP}}
    \end{array}
\right.
\end{align}
which is normalized so that the average energy per \gls{FBS} is one.
The baseband spectrum of the transmitted signal (with \gls{CP}) is therefore
\begin{align}
    \label{eq:s_cp}
    S_{m}^{\text{CP}}(f) = \sum_{n \in \mathcal{N}} x_{m, n} \frac{1}{\sqrt{\Delta_f^{\text{CP}}}}\text{sinc} \left( \frac{f-n\Delta_f}{\Delta_f^{\text{CP}}} \right)
\end{align}
where $\Delta_f^{\text{CP}} = \frac{1}{T+T^{\text{CP}}}$.

\subsubsection{Relevant metrics}
\Gls{OFDM} waveforms have, inter alia, two major drawbacks.
The first one is their high amplitude peaks, which create distortions in the output signal due to \gls{PA} saturation.
Such distortions are usually reduced by operating the PA with a large power back-off or by leveraging complex digital pre-distortion, thus reducing the power efficiency.
Let us denote by $\alpha(t) = \frac{|s(t)|^2}{\EE \left[|s(t)|^2 \right]} $ the ratio between the instantaneous and average power of a signal.
We define the $\text{PAPR}_{\epsilon}$ as the smallest $e\geq 0$, such that the probability of $\alpha$ being larger than $e$ is smaller than a threshold $\epsilon \in \left( 0, 1 \right)$:
\begin{align}
    \label{eq:papr}
    \text{PAPR}_{\epsilon}  \coloneqq \mathrm{min} \; e, \;\; \text{s. t.} \;\;  P\left(\alpha(t) > e  \right) \leq \epsilon.
\end{align}
Setting $\epsilon=0$ leads to the more conventional \gls{PAPR} definition $\frac{\max{|s(t)|^2}}{\EE \left[|s(t)|^2 \right]}$.
However, the maximum signal power occurs with very low probability, and therefore such a definition of the \gls{PAPR} only has a limited practical interest. 
Relaxing $\epsilon$ to values greater than 0 allows considering more frequent, and therefore more practically relevant, power peaks.

\begin{figure}
    \centering
\begin{tikzpicture}

\definecolor{color0}{rgb}{0.12156862745098,0.466666666666667,0.705882352941177}

\pgfplotsset{
    width=.4\textwidth,
    height=0.35\textwidth
}

\begin{axis}[
tick align=outside,
tick pos=left,
x grid style={white!69.0196078431373!black},
xlabel={$\frac{T+T^{\text{CP}}}{T}$},
xmajorgrids,
xmin=0.995, xmax=1.105,
xtick style={color=black},
y grid style={white!69.0196078431373!black},
ylabel={$E_{I_m}$},
scaled y ticks=false,
yticklabel=\pgfkeys{/pgf/number format/.cd,fixed,precision=3,zerofill}\pgfmathprintnumber{\tick},
ymajorgrids,
ymin=24.8068725948848, ymax=24.8139566514653,
ytick style={color=black}
]
\addplot [semithick, color0, mark=+, mark size=3, mark options={solid}]
table {%
1 24.8071945974566
1.01 24.8074462917281
1.02 24.8079299281401
1.03 24.8086259535118
1.04 24.8093693671574
1.05 24.8100782735769
1.06 24.8107878318214
1.07 24.8115202041498
1.08 24.8122389328379
1.09 24.8129317426784
1.1 24.8136346488935
};
\end{axis}

\end{tikzpicture}
\caption{Effect of the \gls{CP} length on the in-band energy.}
\label{fig:cp}
\end{figure}
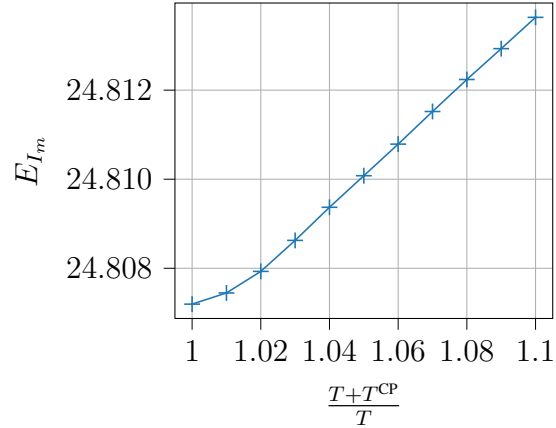

The second drawback of OFDM is its low spectral containment. 
This characteristic is typically measured with the \gls{ACLR}, which is the ratio between the expected out-of-band energy $\EE_{\xv_m}\left[ E_{O_m} \right]$ and the expected in-band energy $\EE_{\xv_m}\left[ E_{I_m}\right]$:
\begin{align}
    \label{eq:aclr}
    \text{ACLR} \coloneqq  \frac{\EE_{\xv_m} \left[ E_{O_m} \right]}{\EE_{\xv_m} \left[ E_{I_ m}\right]} 
     =   \frac{\EE_{\xv_m} \left[ E_{A_m} \right]}{\EE_{\xv_m} \left[ E_{I_m} \right]}-1 
\end{align}
where $E_{O_m}$, $E_{I_m}$, and $E_{A_m} = E_{O_m} + E_{I_m}$ are respectively the out-of-band, in-band, and total energy of the OFDM symbol $m$.
The in-band energy $E_{I_m}$ is given by
\begin{equation}
\begin{split}
E_{I_m} & \coloneqq \int_{-\frac{N \Delta_f}{2}}^{\frac{N \Delta_f}{2}} \left| S_m (f)\right|^2 df = \xv_m ^H \Vm \xv_m\\
\end{split} 
\end{equation}
where each element $v_{a, b}$ of the matrix $\Vm \in \RR^{N \times N}$ is 
\begin{equation}
v_{a, b} = \frac{1}{\Delta_f^{\text{CP}}} \int_{-\frac{N \Delta_f}{2}}^{\frac{N \Delta_f}{2}}  \text{sinc} \left(\frac{f - a \Delta_f}{\Delta_f^{\text{CP}}} \right) \text{sinc} \left(\frac{f - b \Delta_f}{\Delta_f^{\text{CP}}} \right) df, \quad a, b \in \Nc.
\end{equation}
The effect on the \gls{CP} length on the in-band energy is shown in~Fig.\ref{fig:cp}, which have been obtained by sending $10^6$ random \glspl{FBS} $\xv_m \thicksim \Cc \Nc (\mathbf{0}, \frac{1}{\sqrt{2}}\Id)$ on $N=25$ subcarriers with \gls{CP} lengths of  $T^{\text{CP}} \in [0, 0.1T]$.
It can be seen that the in-band energy increases with the CP length, which is to be expected as increasing  $T^{\text{CP}}$ amounts to modulating the subcarriers with sinc for which the ripples are brought closer together, thus containing more energy in $\LSB \Delta_f \LB n-\frac{1}{2} \RB, \Delta_f \LB n+\frac{1}{2} \RB \RSB$.
This in-band energy increase can be directly mapped to an \gls{ACLR} decrease, as the total energy does not depend on CP length.
In the following, we therefore consider that $T^{\text{CP}} = 0$ when computing the time-domain representation and spectrum of the signal, as it corresponds to the worst-case scenario in terms of spectral energy leakage. 

Finally, the total energy can be more conveniently computed in the time domain:
\begin{align}
E_{A_m} \coloneqq \int_{ -\frac{T}{2}}^{\frac{T}{2}} \left| s (t)\right|^2 dt = \xv_m ^H \Wm \xv_m
\end{align}
where $\Wm \in \RR^{N \times N}$ has elements
\begin{equation}
w_{a, b} = \frac{1}{T} \int_{-\frac{T}{2}}^{\frac{T}{2}}  e^{i 2 \pi (a-b) t /T} dt, \quad a, b \in \Nc .
\end{equation}

\subsection{Baseline}
One technique to reduce the \gls{PAPR} of OFDM signal is \gls{TR}, in which a subset of $R$ tones (subcarriers) is used to create peak-reduction signals.
These subcarriers are referred to as \glspl{PRT}, and the remaining $D$ subcarriers are used for data and pilot transmission.
The sets containing the \glspl{PRT} and the data-carrying subcarriers are respectively denoted by $\mathcal{R}$ and $\mathcal{D}$, and are such that $\mathcal{R} \cup \mathcal{D} = \mathcal{N}$.

\subsubsection{Transmitter}
The \gls{TR}-based transmitter sends three types of signals: data signals, peak reduction signals, and pilot signals which are used by the receiver to estimate the channel.
Such pilots are inserted in the \gls{RG} following the 5G \gls{NR} pattern visible in Fig.~\ref{fig:pilots}, i.e., every two \glspl{RE} on the second \gls{OFDM} symbol, and the value of each pilot is chosen randomly on the unit circle. 
For clarity, only data and peak-reduction signals are considered when describing the transmitter, as transmitting pilots is achieved by simply replacing some \glspl{RE} carrying data by reference signals.
We denote by $u_{m,n\in\mathcal{D}}$ and $c_{m,n\in\mathcal{R}}$ the \glspl{FBS} carrying data and peak-reduction signals, respectively.
An \gls{FBS} $ u_{m, n\in\mathcal{D}}$ corresponds to the mapping of a vector of bits $\bv_{m, n\in\mathcal{D}}$ following a $2^K$-\gls{QAM}, the constellation of which is denoted by $\mathcal{C}\in \CC^{2^{K}}$.
The vector of \glspl{FBS} $\uv_{m}\in\CC^{N}$ is composed of all $ u_{m, n\in\mathcal{D}}$ and of zeros at positions that correspond to  \glspl{PRT}, i.e., $u_{m, n\in\mathcal{R}} = 0$.
The reduction vector $\cv_{m}\in\CC^{N}$ is formed by the signals $c_{m,n\in\mathcal{R}}$ mapped to the \glspl{PRT}, and is conversely such that $c_{m, n\in\mathcal{D}} = 0$.
As an example, if three subcarriers are used and only the last one is used as a \gls{PRT}, these vectors are expressed as $\uv_m = \left[ u_{m, -1}, u_{m, 0}, 0 \right]\tp$ and $\cv_m = \left[ 0, 0, c_{m, 1} \right]\tp$.
The vector of discrete baseband signals to be transmitted and the corresponding continuous-time waveform are finally denoted by $ \xv_m = \uv_m + \cv_m$ and $s_m(t)$, respectively.
\gls{TR} aims at finding $\cv_m$ that minimizes the maximum squared signal amplitude, i.e., 
\begin{align}
    \arg\min_{\cv_m} \max_{t} |s_m (t)|^2.
    \label{eq:papr_1}
\end{align}
As minimization over the time-continuous signal leads to intractable calculations, $s_m (t)$ is first discretized. 
Many previous studies considered a discrete vector $\zv_m \in \CC^{N}$, sampled with a period $\frac{T}{N}$, as a substitute for the underlying signal~\cite{wang2021model, 8928103, 8240644}.
However, it has been shown that using a vector $\underline{\zv}_m \in \CC^{NO_s}$, oversampled by a factor $O_s$, is necessary to correctly represent the analog waveform~\cite{922754}.
\begin{figure}
    \centering
\input{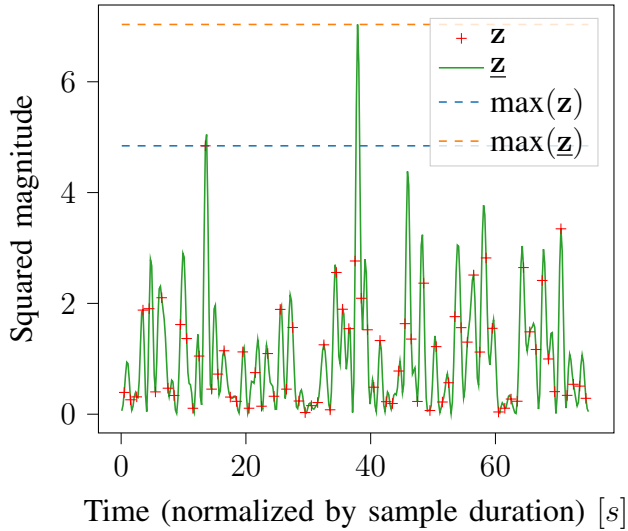}
\caption{OFDM signal generated from $N=75$ subcarriers.}
\label{fig:waveform}
\end{figure}
The difference between the two discretized vectors are visible in Fig.~\ref{fig:waveform}, where the squared amplitude of $\zv_m$ and $\underline{\zv}_m$ are plotted for an arbitrary OFDM symbol, with $N=75$ subcarriers and an oversampling factor of $O_s=5$.
It can be seen that the oversampled signal exhibits a different maximum peak with a higher amplitude as well as numerous secondary peaks that are not present in the non-oversampled signal.
These peaks might lie in the \gls{PA} saturation region, causing distortions of the transmitted waveform.
Let us define the \gls{IDFT} matrix $\Fm\htp \in \CC^{NO_s \times N}$, where each element is expressed as $f_{a, b} = \frac{1}{\sqrt{N}O_s}e^{\frac{j2\pi a b}{NO_s}}$.
The oversampled vector can be obtained with
\begin{align}
    \underline{\zv}_m = \Fm\htp \xv_m = \Fm \left( \uv_m + \cv_m \right).
\end{align}

The value of $\cv_m$ that minimizes the \gls{PAPR} can now be approximately found by minimizing the oversampled signal:
\begin{align}
    \arg\min_{\cv_m} \left\Vert g \left( \Fm\htp (\uv_m + \cv_m) \right) \right\Vert_{\infty}
\label{eq:papr_2}
\end{align}
where $g(\cdot)$ denotes the element-wise squared magnitude $| \cdot |^{2}$ and $ \left\Vert \cdot \right\Vert_{\infty}$ denotes the infinity norm.
\sloppy The convexity of $\left\Vert g \left( \Fm (\uv_m + \cv_m) \right) \right\Vert_{\infty}$ theoretically allows to find the optimal value of $\cv_m$, but the associated complexity leads to the development of algorithms that approximate this value in a limited number of iterations~\cite{1261335}.
In this work, however, we use a convex solver~\cite{diamond2016cvxpy} to find the exact solution of \eqref{eq:papr_2} for each symbol $\xv_m$.
Although such a scheme would be prohibitively complex in practice, it is considered here to provide a close to ideal implementation of a TR-based baseline.
Moreover, for fairness with conventional \gls{QAM} systems, we add the convex constraint $\cv_m \htp \cv_m \leq R$ so that the average energy per OFDM symbol equals at most $N$. 
We experimentally verified that the average energy of the peak reduction signals $\EE_{\cv_m}\left[ \cv_m\htp \cv_m \right] $ was always close to $R$, leading to $ \EE_{\xv_m}\left[ \xv_m\htp \xv_m \right] \approx N$.
Finally, it was shown that placing the \glspl{PRT} at random locations at every transmission leads to the lowest \gls{PAPR} among other placement techniques~\cite{fcd29097-44f6-450e-8fc9-981cffc60574}.
The baseline therefore implements such a random positioning scheme for all OFDM symbols, except for the one carrying pilots for which peak-reduction signals cannot be inserted on pilot-carrying subcarriers.
On this specific OFDM symbol, the number of \glspl{PRT} is also reduced to $\frac{R}{2}$ in order to always maintain a significant number of subcarriers carrying data.

\subsubsection{Receiver}
On the receiver side, channel estimation is performed first, using the pilot signals received in the pilot-carrying OFDM symbol $m^{(p)}\in\mathcal{M}$.
The pattern depicted in Fig.~\ref{fig:pilots} allocates $\frac{N+1}{2}$ \glspl{RE} to pilot transmissions. 
Let us denote by $\pv_{m^{(p)}}\in\CC^{\frac{N+1}{2}}$ the vector of received pilot signals, extracted from $\yv_{m^{(p)}}$.
The channel covariance matrix, providing the spectral correlations between all \glspl{RE} carrying pilots, is denoted by $\Sigmam\in\CC^{\frac{N+1}{2} \times \frac{N+1}{2}}$.
This matrix can be empirically estimated by constructing a large dataset of received pilot signals and computing the statistics over the entire dataset.
The channel coefficients at \glspl{RE} carrying pilots are estimated using \gls{LMMSE}:
\begin{align}
    \widehat{\hv}_{m^{(p)}}^{(p)} = \Sigmam \left( \Sigmam + \sigma^2 \mathbf{I}_{\frac{N+1}{2}} \right)^{-1} \pv_{m^{(p)}} \quad \in \CC^{\frac{N+1}{2}} .
\end{align}
Channel estimation at the remaining $N$ \glspl{RE} of the OFDM symbol $m^{(p)}$ is achieved through linear interpolation.
As the channel is assumed to be invariant over the duration of a slot, the so obtained vector $\widehat{\hv}_{m^{(p)}}  \in \CC^{N}$ is also used for all other OFDM symbols, forming the channel estimate matrix $\widehat{\Hm} \in \CC^{M \times N}$ where all columns are equal.
On fast changing channels, pilots could be inserted in other OFDM symbols to better track the evolution of the channel.

The transmitted \glspl{FBS} are estimated through equalization:
\begin{align}
    \widehat{\Xm} = \Ym \oslash \widehat{\Hm}.
\end{align}
Finally, the \gls{LLR} of the $k^{\text{th}}$ bit corresponding to the RE $(m,n)$ is computed with a conventional \gls{AWGN} demapper:
\begin{align}
    \text{LLR}_{m,n}(k) = \text{ln} \left( \frac{
        \sum_{c \in \mathcal{C}_{k, 1}} \text{exp} \left( - \frac{|\hat{h}_{m,n}|^2 }{\sigma^2 } | \hat{x}_{m,n} - c |^2 \right)}
        {\sum_{c \in \mathcal{C}_{k, 0}} \text{exp} \left( - \frac{|\hat{h}_{m,n}|^2 }{\sigma^2 } | \hat{x}_{m,n} - c |^2 \right)} 
        \right)
\end{align}
where $\mathcal{C}_{k, 1}$ ($\mathcal{C}_{k, 0}$) is the subset of $\mathcal{C}$ containing the symbols that have the $k^{\text{th}}$ bit set to 1 (0), and $\frac{|\hat{h}_{m,n}|^2 }{\sigma^2 }$ is the post-equalization noise variance.

\section{Learning a high dimensional modulation} 
\label{sec:hig_dimensional_modulation}

\begin{figure}[t]
    \centering
    \includegraphics[width=0.50\textwidth]{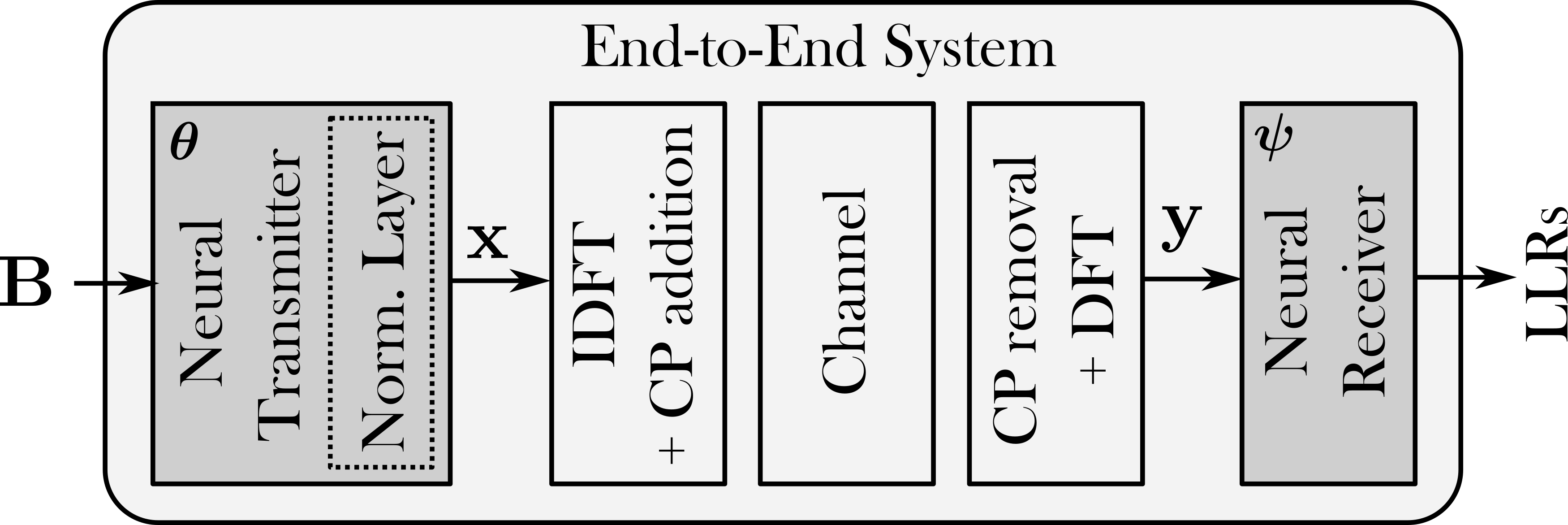}
    \caption{Trainable system, where grayed blocks represent trainable components.}
    \label{fig:E2E_system}
\end{figure}

In the following, we train an \gls{NN}-based transmitter and receiver to maximize an achievable rate under \gls{ACLR} and \gls{PAPR} constraints.
This end-to-end system is referred to as "E2E" system for brevity, and is schematically shown in Fig.~\ref{fig:E2E_system}.
An optimization procedure is derived to handle the constrained optimization problem, in which the loss function is expressed as a differentiable augmented Lagrangian. 
Next, we detail the transmitter and receiver architectures, both implemented as \glspl{CNN}.

\subsection{Optimization procedure}

\subsubsection{Problem formulation}
We aim at finding a high-dimensional modulation and associated detector that both maximize an information rate for the OFDM transmission and satisfy constraints on the signal \gls{PAPR} and \gls{ACLR}.
The transmitter and receiver of the E2E system operate on top of OFDM, i.e., IDFT (DFT) is performed and a cyclic prefix is added (removed) before (after) transmission (see Fig.~\ref{fig:E2E_system}).
The considered rate \cite{pilotless20} depends on the transmitter and receiver trainable parameters, respectively denoted by $\thetav$ and $\psiv$, and is achievable assuming a bit metric decoder~\cite{bocherer2017achievable}:
\begin{align}
    \label{eq:rate0}
    C(\thetav, \psiv) & = \frac{1}{MN}\sum_{m \in \mathcal{M}}\sum_{n \in \mathcal{N}}\sum_{k=0}^{K-1}I\left( b_{m,n,k}; \yv_m \right | \thetav)\\
    & - \frac{1}{MN}\sum_{m\in\mathcal{M}}\sum_{n \in \mathcal{N}}\sum_{k=0}^{K-1}\EE_{\yv_m}\left[\text{D}_\text{KL} \left( P (b_{m,n,k}|\yv_m)|| \widehat{P}_{\psiv}(b_{m,n,k}| \yv_m ) \right)\right]. \nonumber
\end{align}
The first term in \eqref{eq:rate0} is the sum of the mutual informations between all $b_{m,n,k}$ and $\yv_m$, and corresponds to an achievable information rate assuming an ideal receiver that computes the true posterior distributions $P(b_{m,n,k}|\yv_m)$.
The second term is the sum of the expected values of the Kullback–Leibler divergences between the estimated posterior probabilities $ \widehat{P}_{\psiv}(b_{m,n,k}| \yv_m ) $ and the true ones $ P (b_{m,n,k}|\yv_m)$, and corresponds to a rate loss due to the suboptimal receiver \cite{pilotless20}.
It can be seen as a measure of distance between the probabilities that would be computed by an ideal receiver and the ones estimated by our NN-based implementation.
As the E2E system directly outputs \glspl{LLR}, the estimated posterior probabilities can be obtained form
\begin{equation}
    \text{LLR}_{m,n}(k) \coloneqq \text{ln} \left( \frac{\widehat{P}_{\psiv}(b_{m,n,k} = 1| \yv_m ) }{\widehat{P}_{\psiv}(b_{m,n,k} = 0| \yv_m ) } \right).
\end{equation}

To ensure a unit average energy per \gls{RE}, a normalization layer is added to the transmitter (see  Fig.~\ref{fig:E2E_system}).
Perfect normalization would perform
\begin{align}
    \label{eq:layer_norm*}
    l_{\text{norm}}^*(\xv_m) = \frac{\xv_m}{ \left( \frac{1}{N} \EE_{\xv_m}[E_{A_m}] \right)^{\frac{1}{2}} }
\end{align}
but the $2^{MNK}$ different combinations of bits would make the computation of the expected value too complex for any practical system.
Batch normalization is therefore preferred, ensuring that the average energy per RE in the batch is one:
\begin{align}
    \label{eq:layer_norm}
    l_{\text{norm}}(\xv_m^{[j]}) = \frac{\xv_m^{[j]}}{  \left( \frac{1}{M N B_s}  \sum_{m\in\mathcal{M}}\sum_{i=1}^{B_s}  \xv_m^{[i]^{\mathsf{H}}} \Wm \xv_m^{[i]} \right)^{\frac{1}{2}} }
\end{align}
where the superscript $[j]$ denotes the $j^{\text{th}}$ element in the batch.
This expression is slightly different from the one typically used in related works, since it accounts for the correlation that can appear between the \glspl{FBS} generated by the transmitter.
Conventional bit-interleaved modulation systems produces \gls{iid} symbols, and therefore does not need to take such correlation into account.

We can now formulate the constrained optimization problem we aim to solve: 
\begin{subequations}
    \label{eq:rate}
     \begin{align}
    \underset{\thetav, \psiv}{\text{maximize}} & \quad\quad C(\thetav, \psiv) \label{eq:rate1} \\
   \text{subject to}  & \quad\quad \text{PAPR}_{\epsilon}(\thetav) = \gamma_{\text{peak}} \label{eq:rate3} \\
    &  \quad\quad \text{ACLR}(\thetav) \leq \beta_{\text{leak}} \label{eq:rate4} 
     \end{align}
\end{subequations}
where $\gpeak$ and $\bleak$ respectively denote the target \gls{PAPR} and \gls{ACLR}.
Note that the \gls{PAPR} and \gls{ACLR} depend on the transmitter parameters $\thetav$.

\subsubsection{System training}
One of the main advantages of implementing the transmitter-receiver pair as an E2E system is that it enables optimization of the trainable parameters through \gls{SGD}.
This requires a differentiable loss function so that the gradients can be computed and backpropagated through the E2E system.
In the following, the augmented Lagrangian method is leveraged to convert the problem \eqref{eq:rate} into its augmented Lagrangian, which acts a differentiable loss function that can be minimized with respect to $\thetav$ and $\psiv$~\cite{bertsekas2014constrained}.
The key idea is to relax the constrained optimization problem into a sequence of unconstrained problems that are solved iteratively.
This method is known to be more effective than the quadratic penalty method, enabling a faster and more stable convergence~\cite{nocedal2006numerical}.
In the following, we express the objective \eqref{eq:rate1} and the constraints \eqref{eq:rate3} and \eqref{eq:rate4} as differentiable functions that can be evaluated during training and minimized with \gls{SGD}.

First, the achievable rate \eqref{eq:rate1}  can be equivalently expressed using the system \gls{BCE}~\cite{9118963}, which is widely used in binary classification problems:
\begin{align}
    \label{eq:CE}
    L_C(\thetav, \psiv) &:= - \frac{1}{MN}\sum_{m\in\mathcal{M}}\sum_{n\in\mathcal{N}}\sum_{k=0}^{K-1} \EE_{\yv_m} \left[ \text{log}_2 \left(\widehat{P}_{\psiv} (b_{m,n,k}| \yv_m ) \right) \right]  \\
    & = K -  C(\thetav, \psiv). 
\end{align}
To overcome the complexity associated with the computation of the expected value, an approximation is typically obtained through Monte Carlo sampling:
\begin{align}
    \label{eq:CE_batch}
    L_C(\thetav, \psiv) \approx & - \frac{1}{MNB_s}\sum_{m\in\mathcal{M}}\sum_{n\in\mathcal{N}}\sum_{k=0}^{K-1}\sum_{i=0}^{B_s-1}  \text{log}_2 \left( \widehat{P}_{\psiv} \left( b_{m,n,k}^{[i]}| \yv_m ^{[i]} \right) \right).
\end{align}

Second, evaluating the constraint \eqref{eq:rate3} requires the computation of the probability $P(\frac{|s(t)|^2}{\EE \left[|s(t)|^2 \right]}  > e)$, where $e$ is the energy threshold defined in \eqref{eq:papr}.
However, computing such probability would be prohibitively complex due to the sheer amount of possible OFDM symbols.
During training, we therefore enforce the constraint by setting $\epsilon=0$ and penalizing all signals whose squared amplitude exceed $\gpeak$. 
With $\epsilon=0$, the constraint \eqref{eq:rate3} is equivalent to enforcing $L_{\gamma_{\text{peak}}}(\thetav) =0$, with
\begin{align}
    \label{eq:loss_papr}
    L_{\gamma_{\text{peak}}}(\thetav)  = \EE_m \left[ \int_{-\frac{T}{2}}^{\frac{T}{2}} \left(|s_m(t)|^2-\gamma_{\text{peak}} \right)^+ dt \right]
\end{align}
where $(x)^+$ denotes the positive part of $x$, i.e., $ (x)^+= \text{max}(0, x)$.
To evaluate $L_{\gamma_{\text{peak}}}(\thetav)$ during training, the value of the expectation can be obtained through Monte Carlo sampling, and the integral can be approximated using a Riemann sum:
\begin{align}
    L_{\gamma_{\text{peak}}}(\thetav)  \approx \frac{T}{B_s N O} \sum_{i=0}^{B_s-1} \sum_{t=-\frac{NO_s-1}{2}}^{\frac{NO_s-1}{2}}  \left(\left| \underline{z}_{m,t}^{[i]} \right| ^2 - \gamma_{\text{peak}} \right)^+
    \label{eq:loss_papr2}
\end{align}
where $\underline{\zv}_m = \Fm^{-1}\xv_m \in \CC^{NO_s}$ is the vector of the oversampled time signal corresponding to the neural transmitter output  $\xv_m$.

Third, the inequality constraint \eqref{eq:rate4} can be converted to the  equality constraint $\text{ACLR}(\thetav) - \beta_{\text{leak}}  = -q$, where $q\in\RR_+$ is a positive slack variable.
This equality constraint is then enforced by minimizing $L_{\beta_{\text{leak}}}(\thetav) + q $, with
\begin{align}
L_{\beta_{\text{leak}}}(\thetav)  & =  \frac{ \EE \left[ E_A \right]}{ \EE \left[ E_I \right]}-1   - \beta_{\text{leak}} \\
& \approx \frac{  \frac{1}{B_s} \sum_{i=0}^{B_s-1}  \xv^{[i]^{\mathsf{H}}} \Wm \xv^{[i]}}{ \frac{1}{B_s} \sum_{i=0}^{B_s-1}  \xv^{[i]^{\mathsf{H}}} \Vm \xv^{[i]}} -1   - \beta_{\text{leak}} .
\end{align}

Finally, for $\epsilon=0$, the problem \eqref{eq:rate} can be reformulated as 
\begin{subequations}
    \label{eq:pb}
     \begin{align}
    \underset{\thetav, \psiv}{\text{minimize}} & \quad\quad L_C(\thetav, \psiv) \label{eq:pb1} \\
    \text{subject to} & \quad\quad L_{\gamma_{\text{peak}}}(\thetav)  = 0 \label{eq:pb2} \\
    &  \quad\quad L_{\beta_{\text{leak}}}(\thetav) + q = 0 \label{eq:pb3} 
     \end{align}
\end{subequations}
where the objective and the constraints are differentiable and can be estimated at training.
The augmented Lagrangian method introduces two types of hyperparameters that are iteratively updated during training.
The first one corresponds to the penalty parameters which are slowly increased to penalize the constraint with increasing severity. 
The second one corresponds to estimates of the Lagrange multipliers, as defined in~\cite{bertsekas2014constrained}. 
Let us denote by  $\mu_p > 0$ and $\mu_l>0$ the penalty parameters and by $\lambda_p$ and $ \lambda_l$ the Lagrange multipliers for the constraint functions $L_{\gamma_{\text{peak}}}(\thetav)$ and $L_{\beta_{\text{leak}}}(\thetav)$, respectively.
The corresponding augmented Lagrangian is defined as~\cite{bertsekas2014constrained}
\begin{align}
    \label{eq:lagrange_slack}
    \overline{L}^* (\thetav, \psiv, \lambda_p, \lambda_l, & \mu_p, \mu_l, q)  = L_C(\thetav, \psiv) \nonumber \\
    & + \lambda_p L_{\gamma_{\text{peak}}}(\thetav) + \frac{1}{2} \mu_p |L_{\gamma_{\text{peak}}}(\thetav)|^2 \\
    & + \lambda_l  \left( L_{\beta_{\text{leak}}}(\thetav) + q \right) + \frac{1}{2} \mu_l \left\lvert L_{\beta_{\text{leak}}}(\thetav) + q \right\rvert^2. \nonumber
\end{align}
As derived in~\cite{bertsekas2014constrained}, the minimization of~\eqref{eq:lagrange_slack} with respect to $q$ can be carried out explicitly for each fixed pair of $\{\thetav, \psiv\}$ so that the augmented Lagrangian can be equivalently expressed as
\begin{align}
    \label{eq:lagrange}
    \overline{L} (\thetav, \psiv, \lambda_p, \lambda_l, & \mu_p, \mu_l)  = L_C(\thetav, \psiv) \nonumber \\
    & + \lambda_p L_{\gamma_{\text{peak}}}(\thetav) + \frac{1}{2} \mu_p |L_{\gamma_{\text{peak}}}(\thetav)|^2 \\
    & + \frac{1}{2\mu_l} \left( \text{max}(0, \lambda_l + \mu_l L_{\beta_{\text{leak}}}(\thetav) )^2 - \lambda_l^2 \right). \nonumber
\end{align}
Each training iteration comprises multiples steps of SGD on the augmented Lagrangian \eqref{eq:lagrange} followed by an update of the hyperparameters.
The optimization procedure is detailed in Algorithm \ref{alg:lagrangian}, where $\tau \in \RR^+$ controls the evolution of the penalty parameters and the superscript $(u)$ refers to the $u^{\text{th}}$ iteration of the algorithm.

\begin{algorithm}
    \SetAlgoLined
     Initialize $\thetav, \psiv, \lambda_p^{(0)}, \lambda_l^{(0)}, \mu_p^{(0)}, \mu_l^{(0)}$ \\
     \For{$u = 0, ... $}{
      $\triangleright$ Perform multiple steps of SGD \\
      on $\overline{L} (\thetav, \psiv, \lambda, \lambda_l, \mu_p, \mu_l)$ w.r.t. $\thetav$ and $ \psiv$ \\
      $\triangleright$ Update optimization hyperparameters: \\
      $\lambda_p^{(u+1)} = \lambda_p^{(u)} + \mu_p^{(u)} L_{\gamma_{\text{peak}}}(\thetav) $\\
      $\lambda_l^{(u+1)} = \text{max} \left(0, \lambda_l^{(u)}  + \mu_l^{(u)} L_{\beta_{\text{leak}}}(\thetav) \right)$\\
      $\mu_p^{(u+1)} = (1+\tau) \mu_p^{(u)}$ \\
      $\mu_l^{(u+1)} = (1+\tau) \mu_l^{(u)}$ 
     }
     \caption{Training procedure}
     \label{alg:lagrangian}
\end{algorithm}

\subsection{System architecture}

\begin{figure}
    \centering
    \begin{subfigure}{.45\textwidth}
        \centering
        \includegraphics[height=170pt]{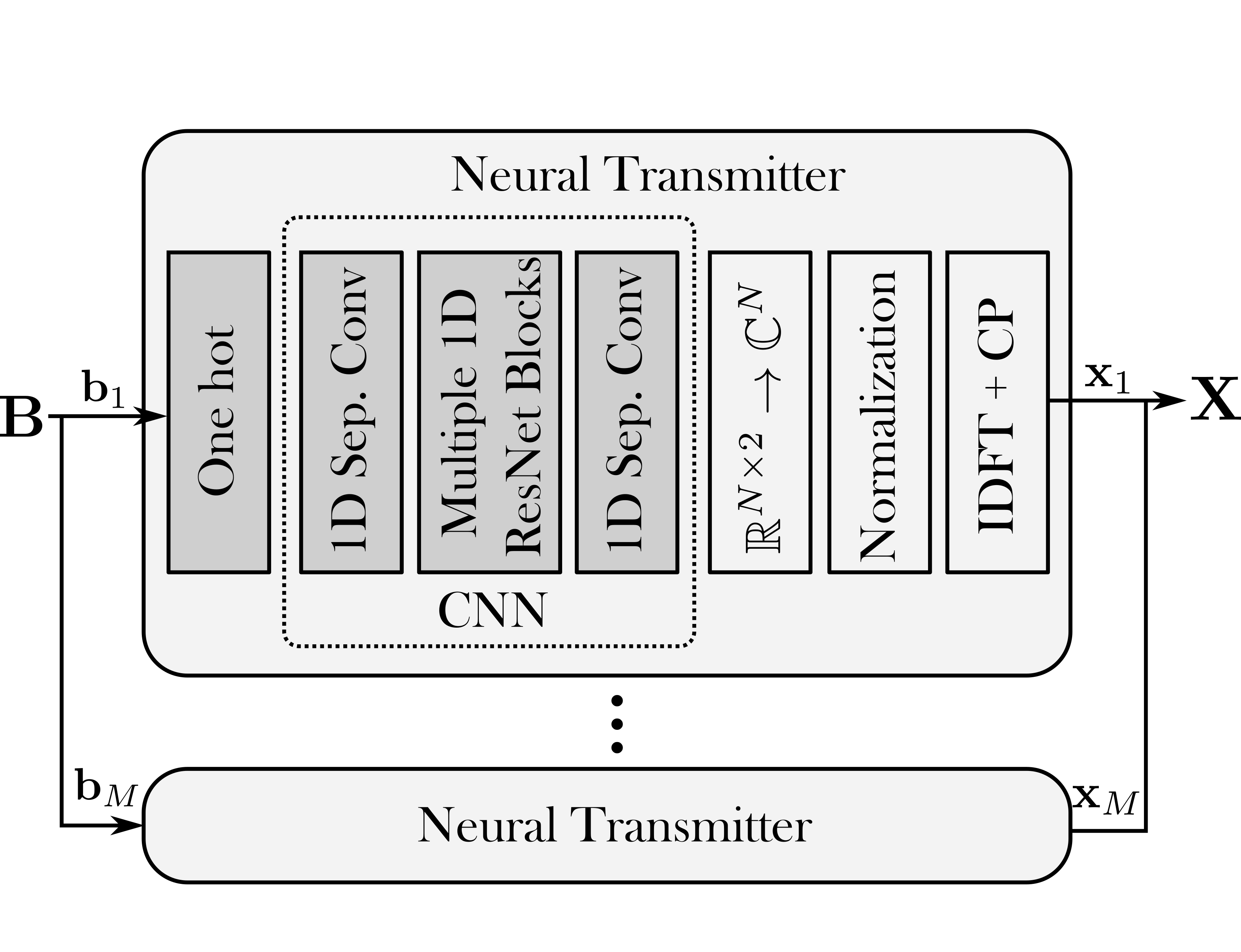}
        \caption{Neural Transmitter.}
        \label{fig:nn_tx}
      \end{subfigure}
      \hfill
      \begin{subfigure}{.2\textwidth}
        \centering
        \includegraphics[height=170pt]{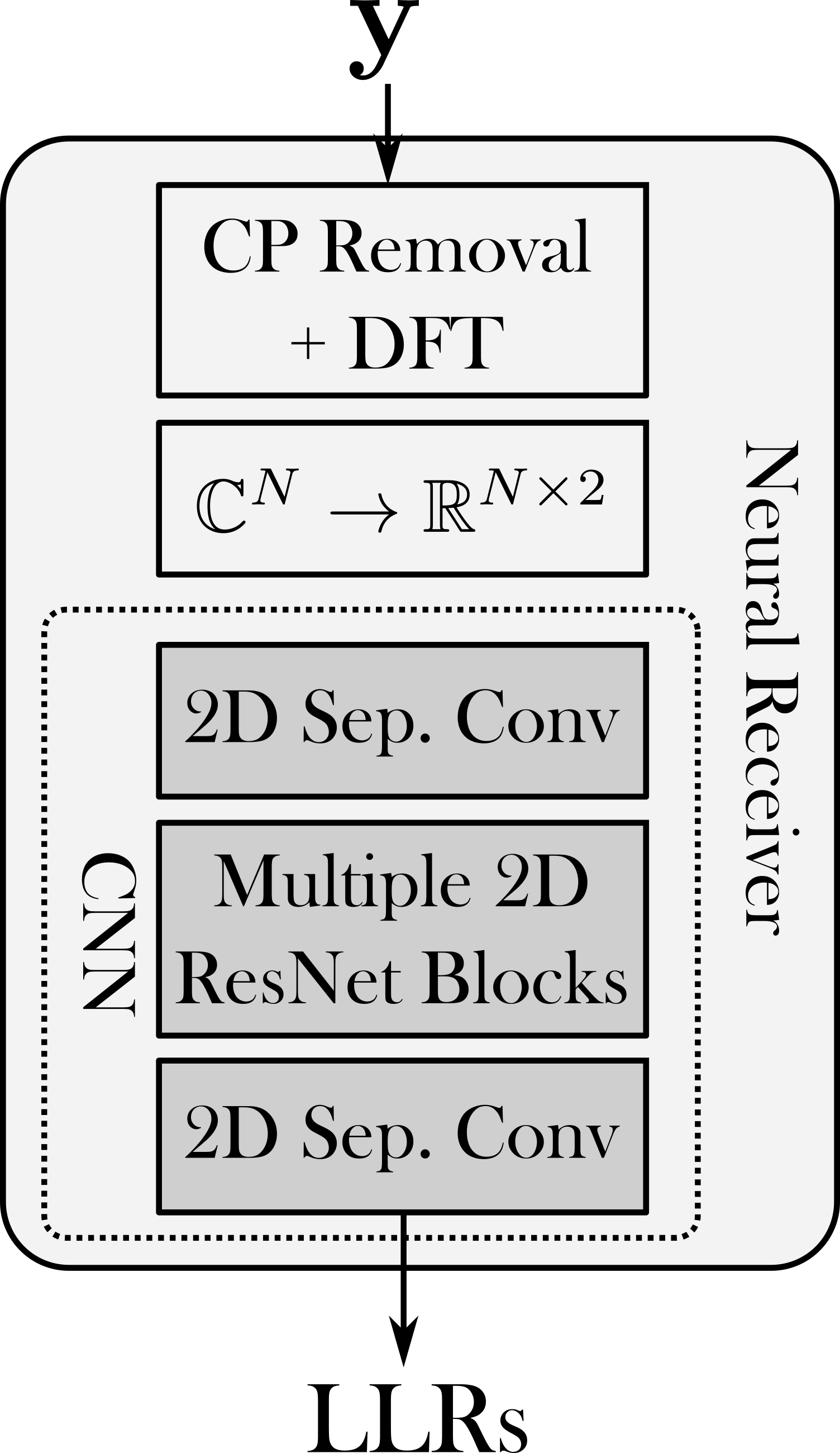}
        \caption{Neural Receiver.}
        \label{fig:nn_rx}
      \end{subfigure}%
      \hfill
    \begin{subfigure}{.2\textwidth}
      \centering
      \includegraphics[height=170pt]{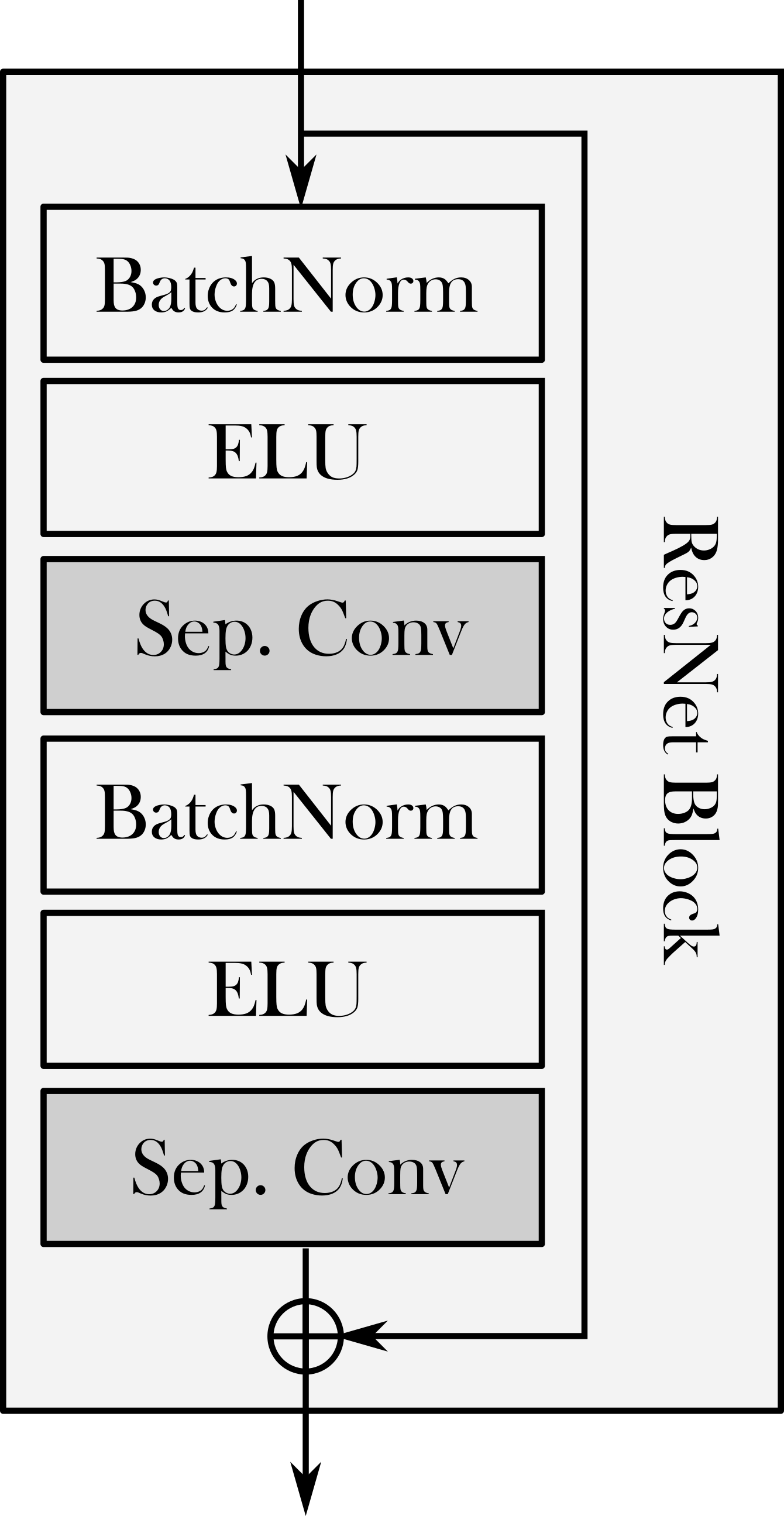}
      \caption{ResNet block.}
      \label{fig:resnet}
    \end{subfigure}%
    
    \caption{Different parts of the end-to-end system, where grayed blocks are trainable components.}
    \label{fig:components}
\end{figure}

The neural transmitter and receiver are based on similar architectures, schematically shown in Fig.~\ref{fig:components}.
The core element is a \emph{ResNet block}, which was introduced in the physical layer to implement a fully \gls{NN}-based radio receiver \cite{honkala2020deeprx}, and whose effectiveness has been demonstrated in other related works~\cite{pilotless20, korpi2020deeprx, goutay2020machine}.
A ResNet block is made of two identical sequences of layers followed by the addition of the input, as depicted in Fig.~\ref{fig:resnet}.
In the original ResNet block~\cite{he2016identity}, each sequence was composed of a batch normalization layer, a rectified linear unit (ReLU) activation function, and a convolution. 
Our architecture differs from the original one in that ReLUs are replaced by exponential linear units (ELUs) to alleviate the vanishing gradient problem~\cite{clevert2016fast} and separable convolutions are used since they enable similar performance than conventional ones but at a fraction of the computational cost~\cite{howard2017mobilenets}.
Finally, we use zero-padding on all 1D (2D) convolutions to maintain constant the size of the first (and possibly second) dimension(s).
In the following, the architectures of the neural transmitter and receiver are detailed, although the exact numbers of parameters for each layer are given in Section~\ref{sec:evaluations}.

The transmitter processes all OFDM symbols in parallel (see Fig.~\ref{fig:nn_tx}). 
Each instance of the CNN implemented at the transmitter takes as input the matrix of bits $\Bm_m$, corresponding to the OFDM symbol $m$, and outputs the OFDM symbol $\xv_m$.
The vector of bits is first converted into its one hot representation, i.e., into vectors of $\{0, 1\}^{2^K}$ where all elements but one are set to zero.
Then, a \gls{CNN} comprises one 1D separable convolution, multiple 1D ResNet blocks, and another 1D separable convolution.
This CNN is fed with the one-hot matrix of dimension $N \times 2^K$, were $N$ corresponds to the dimension of the 1D convolution and $2^K$ to different convolution channels, and outputs $N\times 2$ elements.
The next layers convert these $N\times 2$ real numbers into $N$ complex symbols and normalize them, as in \eqref{eq:layer_norm}, to have a unit average energy per \gls{RE}.
Finally, an \gls{IDFT} is performed on the symbols and a \gls{CP} is added before transmission.
We experimentally verified that independent processing of all OFDM symbols, resulting in the use of 1D convolutions, leads to better performance than 2D convolutions that would process all OFDM symbols at once.
This could be explained by the 1D nature of \gls{PAPR} and \gls{ACLR} measurements, which are computed for all OFDM symbols separately.

The neural receiver, on the contrary, performs a 2D processing on all OFDM symbols since it enables more accurate channel estimation and equalization~\cite{pilotless20, goutay2020machine}.
At reception, the \gls{CP} is first removed and an DFT is applied to the received signals $\Ym$.
The $M \times N$ symbols are then converted into $M \times N \times 2$ real numbers that are fed, along with the transmission SNR of size $M \times N \times 1$, into a 2D CNN.
The architecture of the receiver CNN is similar to the one of the transmitter, except that 2D separable convolutions are used.
The last 2D separable layer outputs $M \times N \times K$ real numbers that correspond to the \glspl{LLR} of all transmitted symbols, as shown in Fig.~\ref{fig:nn_rx}.
Note that no pilots are used as it was shown in~\cite{pilotless20} that pilotless communication is possible over OFDM channels when  neural receivers are used.

\section{Simulations results and insights} 
\label{sec:evaluations_insights}

In this section, the E2E system is benchmarked against the TR baseline.
We first describe the training and evaluation setup, followed by rate, \gls{BER}, and goodput comparisons.

\begin{table}

    \centering
    
    \newcolumntype{M}[1]{>{\centering\arraybackslash}m{#1}}
    \newcolumntype{N}{@{}m{0pt}@{}}
    
    \renewcommand{\arraystretch}{1.2}
    
    \begin{tabular}{ |M{1.5cm}||M{1.5cm}|M{1.5cm}|M{1.5cm}|M{1.5cm}|M{1.5cm}|M{1.5cm}|M{1.5cm}| N }
        \hline
            & Sep. Conv. 1D (2D) & \multicolumn{5}{c|}{ResNet blocks 1D (2D)} & Sep. Conv 1D (2D) \\
		\hhline{|=|=|=====|=|}
        Kernel size & 1 (1,1) & 3 (3,2)& 9 (9,4)& 15 (15,6)& 9 (9,4)& 3 (3,2)& 1 (1,1)\\
        \hline
        Dilation rate & 1 (1,1)& 1 (1,1)& 2 (2,1) & 4 (4,1)& 2 (2,1)& 1 (1,1)& 1 (1,1)\\
        \hline
        \# Filters & \multicolumn{6}{c|}{128} & $2$ ($K$) \\
        \hline
    \end{tabular}
    
    \caption{Parameters for the neural transmitter (receiver).}
    \label{table:arch}
    \renewcommand{\arraystretch}{1}
    
\end{table}

\subsection{Evaluations}
\label{sec:evaluations}

\subsubsection{Training and evaluation setup}
Separate datasets were used for training and testing, both generated using a mixture of \gls{3GPP}-compliant \gls{UMi} line-of-sight (LOS) and non-LOS models. The channel responses were generated using  QuaDRiGa 2.4.0~\cite{quadriga}, and perfect power control was assumed such that the average channel energy per \gls{RE} was one, i.e., $\EE \left[ |h_{m, n}|^2 \right]= 1$.
$N=75$ subcarriers were considered with $M=14$ OFDM symbols using \glspl{CP} of sufficient lengths so that the channel can be represented by \eqref{eq:OFDM_channel} in the frequency domain.
The center carrier frequency and subcarrier spacing were respectively set to 3.5~\si{GHz} and 30~\si{kHz}, the number of bits per channel use was set to $K=4$, and 16-QAM modulation was used by the baseline.
Coded \gls{BER} comparisons were performed using a standard 5G-compliant low-density parity-check (LDPC) code with length 1024 and rate $r=\frac{1}{2}$.
The baseline was evaluated for $R\in\{0, 2, 4, 8, 16, 32\}$ \glspl{PRT} and the E2E system was trained to achieve \gls{ACLR} targets of $\bleak \in \{-20, -30, -40\}$~\si{dB} and \gls{PAPR} targets of $\gpeak \in \{ 4, 5, 6, 7, 8, 9 \}$~\si{dB}.

The transmitter and receiver architectures that were used are detailed in Table~\ref{table:arch}.
Inspired by~\cite{honkala2020deeprx}, the receptive fields of the \glspl{CNN} were increased using dilations, and the kernel sizes had an increasing, then decreasing number of parameters.
The Lagrange multipliers were initialized to $\lambda^{(0)}_p = \lambda^{(0)}_l = 0$ and $\tau$ was set to $0.004$.
The penalty parameters were initialized to $\mu^{(0)}_p = 10^{-1}$ and $\mu^{(0)}_l = 10^{-3}$, which mirrors the fact that $L_{\gamma_{\text{peak}}}(\thetav)$ is usually two orders of magnitude lower than $L_{\beta_{\text{leak}}}(\thetav)$.
The optimization procedure was composed of 2500 iterations in which 20 SGD steps were performed with a learning rate of $10^{-3}$ and a batch size of $B_s = 100$.
The oversampling factor used to compute $\underline{\zv}$ in \eqref{eq:loss_papr2} was set to $O_s = 5$, as it was shown to be sufficient to correctly represent the underlying analog signal~\cite{1261335}.
Finally, the \gls{SNR} 
\begin{align}
	\text{SNR} =  \frac{\EE_{h_{m,n}} \left[ |h_{m,n}|^2 \right]}{\sigma^2}  =  \frac{1}{\sigma^2}
\end{align}
was chosen randomly in the interval $[10, 30]$~\si{dB} for each \gls{RG} in the batch during training.
The simulations parameters are listed in Table~\ref{table:params}.

\begin{table}[h!]
	\centering
	\renewcommand{\arraystretch}{1.2}
	\begin{tabular}{|p{5.5cm}|c|c|c|c|}
	  \hline
	  Parameters  & Symbol (if any) & Value  \\ 
	  \hhline{|=|=|=|}
	  Number of OFDM symbols & $M$ & 14 \\   \hline
	  Number subcarriers & $N$ & 75   \\   \hline
	  Number of bits per channel use & $K$ & 4 \\   \hline
	  \Gls{PAPR} targets & $\gpeak$ & $\{4, 5, 6, 7, 8, 9\}$~\si{dB} \\   \hline
	  \Gls{ACLR} targets & $\bleak$ & $\{-20, -30, -40\}$~\si{dB} \\   \hline
	  \Glspl{PRT} used by the baseline & R & $\{0, 2, 4, 8, 16, 32\}$\\   \hline
	  Batch size & $B_s$ & 100 \glspl{RG}  \\   \hline
	  Oversampling factor & $O_s$ & 5   \\   \hline
	  Center frequency (subcarrier spacing) & - & \SI{3.5}{\GHz} (\SI{30}{\kHz}) \\   \hline
	  Scenario & - & 3GPP 38.901 UMi LOS + NLOS \\   \hline
	  SNR & - & $[10, 30]$~\si{dB} \\   \hline
	  Code length & - & \SI{1024}{\bit} \\   \hline
	  Code rate& $r$ & $\frac{1}{2}$ \\   \hline
	  Learning rate & - & $10^{-3}$ \\   \hline

	\end{tabular}
	
	\vspace{20pt}
	\caption{Training and evaluation parameters.}
	\label{table:params}
  \end{table}

\subsubsection{Evaluation results}
\begin{figure}[t]
	\centering
\begin{tikzpicture}

\definecolor{color0}{rgb}{0.83921568627451,0.152941176470588,0.156862745098039}
\definecolor{color1}{rgb}{0.12156862745098,0.466666666666667,0.705882352941177}
\definecolor{color2}{rgb}{1,0.498039215686275,0.0549019607843137}
\definecolor{color3}{rgb}{0.172549019607843,0.627450980392157,0.172549019607843}

\pgfplotsset{
    width=.55\textwidth,
    height=0.5\textwidth
}

\begin{axis}[
legend cell align={left},
legend style={
  fill opacity=0.8,
  draw opacity=1,
  text opacity=1,
  at={(0.97,0.03)},
  anchor=south east,
  draw=white!80!black
},
tick align=outside,
tick pos=left,
x grid style={white!69.0196078431373!black},
xlabel={$\text{PAPR}_{10^{-3}}$ [\si{dB}]},
xmajorgrids,
xmin=4, xmax=9,
xtick style={color=black},
y grid style={white!69.0196078431373!black},
ylabel={Average rate per RE [\si{\bit}]},
ymajorgrids,
ymin=2, ymax=3.7,
ytick style={color=black}
]
\addplot [semithick, color3, mark=*, mark size=3, mark options={solid}]
table {%
4.3 2.97
5.2 3.3
5.9 3.48
6.8 3.51
7.6 3.53
8.4 3.54
};
\addlegendentry{E2E system,  $\beta_{\text{leak}}=-20$ \si{dB}}
\draw (axis cs:4.3+-0.35, 2.97+0.03) node[
  scale=0.5,
  anchor=base west,
  text=color3,
  rotate=0.0
]{-23.1};
\draw (axis cs:5.2+-0.35, 3.3+0.03) node[
  scale=0.5,
  anchor=base west,
  text=color3,
  rotate=0.0
]{-22.2};
\draw (axis cs:5.9+-0.35, 3.48+0.03) node[
  scale=0.5,
  anchor=base west,
  text=color3,
  rotate=0.0
]{-21.2};
\draw (axis cs:6.8+-0.35, 3.51+0.03) node[
  scale=0.5,
  anchor=base west,
  text=color3,
  rotate=0.0
]{-20.9};
\draw (axis cs:7.6+-0.35, 3.53+0.03) node[
  scale=0.5,
  anchor=base west,
  text=color3,
  rotate=0.0
]{-20.7};
\draw (axis cs:8.4+-0.35, 3.54+0.03) node[
  scale=0.5,
  anchor=base west,
  text=color3,
  rotate=0.0
]{-20.7};

\addlegendentry{E2E system,  $\beta_{\text{leak}}=-30$ \si{dB}}
\addplot [semithick, color2, mark=triangle*, mark size=3, mark options={solid}]
table {%
4.28 2.83
5.25 3.13
6.0 3.35
6.75 3.42
7.6 3.45
8.6 3.47
};
\draw (axis cs:4.28-0.3, 2.83+0.02) node[
  scale=0.5,
  anchor=base west,
  text=color2,
  rotate=0.0
]{-30.7};
\draw (axis cs:5.25-0.4, 3.13+0.013) node[
  scale=0.5,
  anchor=base west,
  text=color2,
  rotate=0.0
]{-30.6};
\draw (axis cs:6.0-0.4, 3.35+0.013) node[
  scale=0.5,
  anchor=base west,
  text=color2,
  rotate=0.0
]{-30.8};
\draw (axis cs:6.75-0.4, 3.42+0.013) node[
  scale=0.5,
  anchor=base west,
  text=color2,
  rotate=0.0
]{-30.8};
\draw (axis cs:7.6-0.4, 3.45+0.006) node[
  scale=0.5,
  anchor=base west,
  text=color2,
  rotate=0.0
]{-31.5};
\draw (axis cs:8.6-0.45, 3.47+0.01) node[
  scale=0.5,
  anchor=base west,
  text=color2,
  rotate=0.0
]{-31.6};

\addlegendentry{E2E system,  $\beta_{\text{leak}}=-40$ \si{dB}}
\addplot [semithick, color0, mark=diamond*, mark size=3, mark options={solid}]
table {%
4.3 2.67
5.2 3.01
6.0 3.12
6.75 3.23
7.7 3.32
8.7 3.39
};
\draw (axis cs:4.3-0.3, 2.67+0.03) node[
  scale=0.5,
  anchor=base west,
  text=color0,
  rotate=0.0
]{-43.0};
\draw (axis cs:5.2-0.4, 3.01+0.01) node[
  scale=0.5,
  anchor=base west,
  text=color0,
  rotate=0.0
]{-40.9};
\draw (axis cs:6.0-0.4, 3.12+0.01) node[
  scale=0.5,
  anchor=base west,
  text=color0,
  rotate=0.0
]{-41.3};
\draw (axis cs:6.75-0.4, 3.23+0.01) node[
  scale=0.5,
  anchor=base west,
  text=color0,
  rotate=0.0
]{-42.2};
\draw (axis cs:7.7-0.4, 3.32+0.01) node[
  scale=0.5,
  anchor=base west,
  text=color0,
  rotate=0.0
]{-40.3};
\draw (axis cs:8.7-0.5, 3.39-0.01) node[
  scale=0.5,
  anchor=base west,
  text=color0,
  rotate=0.0
]{-40.3};

\addlegendentry{Baseline}
\addplot [semithick, color1, mark=square*, mark size=3, mark options={solid}]
table {%
4.7  1.96
5.1  2.7
6.1  3.1
6.9  3.25
7.75  3.36
8.4 3.45
};
\draw (axis cs:4.7+0.05 , 1.96-0.025) node[
  scale=0.5,
  anchor=base west,
  text=color1,
  rotate=0.0
]{-20.6};
\draw (axis cs:5.1+0.05 , 2.7-0.025) node[
  scale=0.5,
  anchor=base west,
  text=color1,
  rotate=0.0
]{-20.8};
\draw (axis cs:6.1+0.05 , 3.1-0.025) node[
  scale=0.5,
  anchor=base west,
  text=color1,
  rotate=0.0
]{-20.5};
\draw (axis cs:6.9+0.05 , 3.25-0.025) node[
  scale=0.5,
  anchor=base west,
  text=color1,
  rotate=0.0
]{-20.5};
\draw (axis cs:7.75+0.06 , 3.36-0.015) node[
  scale=0.5,
  anchor=base west,
  text=color1,
  rotate=0.0
]{-20.5};
\draw (axis cs:8.4+0.07 , 3.45-0.035) node[
  scale=0.5,
  anchor=base west,
  text=color1,
  rotate=0.0
]{-20.5};

\addplot [semithick, black, mark=square, mark size=3, mark options={solid,fill opacity=0}]
table {%
8.4 3.45
};

\addplot [semithick, black, mark=o, mark size=3, mark options={solid,fill opacity=0}]
table {%
5.9 3.48
};

\end{axis}

\end{tikzpicture}
	\caption{Rates achieved by the compared systems. Numbers near scatter plots indicate the \glspl{ACLR}.}
	\label{fig:rates}
\end{figure}
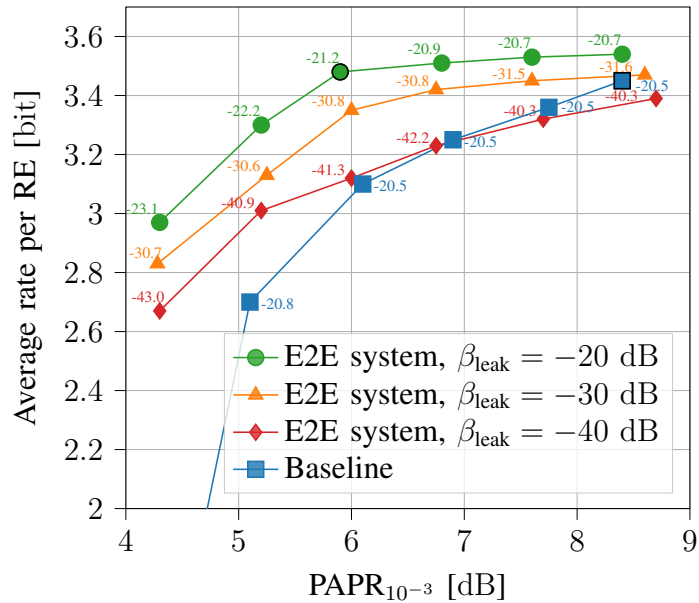

In the following evaluations, the \gls{PAPR} probability threshold was set to $\epsilon = 10^{-3}$.
Note that setting $\epsilon=0$ would only take into account the maximum signal peak, which is achieved by a single possible waveform that has probability $\frac{1}{2^{NK}} \approx 10^{-90}$.
The average rates per \gls{RE} achieved by the baseline and the E2E systems are shown in Fig.~\ref{fig:rates}, where the numbers next to the data points are the corresponding \glspl{ACLR}.
First, it can be seen that at the maximum \gls{PAPR} of approximately $8.5$~\si{dB}, the E2E system trained with $\bleak = -20$ ~\si{dB} achieves a $3\%$ higher throughput than the baseline with no \gls{PRT}. 
This can be explained by the rate loss due to the presence of pilots in the baseline, which do not carry data and account for approximately $4\%$ of the total number of \glspl{RE}.
Second, at lower \glspl{PAPR}, the rates achieved by the E2E system trained with $\bleak\in\{-20, -30\}$ are significantly higher than the ones achieved by the baseline. 
For example, the E2E system trained with $\bleak=-20$ and $\gpeak=5$ achieves an average rate $22\%$ higher than the baseline for the same \gls{PAPR}.  
Finally, the E2E systems are able to meet their respective \gls{PAPR} and \gls{ACLR} targets.

\begin{figure}

	\centering

	\begin{subfigure}[b]{0.2\textwidth}
		\begin{tikzpicture} 
		
			\definecolor{color0}{rgb}{0.12156862745098,0.466666666666667,0.705882352941177}
			\definecolor{color1}{rgb}{1,0.498039215686275,0.0549019607843137}
			\definecolor{color2}{rgb}{0.172549019607843,0.627450980392157,0.172549019607843}
			\definecolor{color3}{rgb}{0.83921568627451,0.152941176470588,0.156862745098039}
		
				\begin{axis}[%
				hide axis,
				xmin=10,
				xmax=50,
				ymin=0,
				ymax=0.4,
				legend columns=4, 
				legend style={draw=white!15!black,legend cell align=left,column sep=1.0ex}
				]
				\addlegendimage{color0, mark=square, mark size=3}
				\addlegendentry{PRT Baseline};
				\end{axis}
		\end{tikzpicture}
	\end{subfigure}
	\hspace{1pt}
	\begin{subfigure}[b]{0.75\textwidth}
	\centering
	\begin{tikzpicture} 
	
	\definecolor{color0}{rgb}{0.12156862745098,0.466666666666667,0.705882352941177}
	\definecolor{color1}{rgb}{1,0.498039215686275,0.0549019607843137}
	\definecolor{color2}{rgb}{0.172549019607843,0.627450980392157,0.172549019607843}
	\definecolor{color3}{rgb}{0.83921568627451,0.152941176470588,0.156862745098039}

    	\begin{axis}[%
    	hide axis,
    	xmin=10,
   	 xmax=50,
    	ymin=0,
    	ymax=0.4,
    	legend columns=4, 
    	legend style={draw=white!15!black,legend cell align=left,column sep=1.0ex}
    	]
    	\addlegendimage{white ,mark=*, mark size=2}
    	\addlegendentry{\hspace{-0.9cm} E2E System with $\bleak=$};
    	\addlegendimage{color2, mark=+, mark size=3}
    	\addlegendentry{$-20$~\si{dB}};
    	\addlegendimage{color1, mark=x, mark size=3}
    	\addlegendentry{$-30$~\si{dB}};
    	\addlegendimage{color3, mark=asterisk, mark size=3}
    	\addlegendentry{$-40$~\si{dB}};
    	\end{axis}
	\end{tikzpicture}
\end{subfigure}

	\vspace{10pt}

	  	\begin{subfigure}[b]{0.45\textwidth}
\begin{tikzpicture}

\definecolor{color0}{rgb}{0.172549019607843,0.627450980392157,0.172549019607843}
\definecolor{color1}{rgb}{1,0.498039215686275,0.0549019607843137}
\definecolor{color2}{rgb}{0.83921568627451,0.152941176470588,0.156862745098039}
\definecolor{color3}{rgb}{0.12156862745098,0.466666666666667,0.705882352941177}

\pgfplotsset{
    width=.88\textwidth,
    height=0.8\textwidth
}

\begin{axis}[
legend cell align={left},
legend style={fill opacity=0.8, draw opacity=1, text opacity=1, draw=white!80!black},
log basis y={10},
tick align=outside,
tick pos=left,
x grid style={white!69.0196078431373!black},
xlabel={$\frac{E_b}{\sigma^2}$},
xmajorgrids,
xtick={10, 12.5, 15, ..., 25},
xmin=9, xmax=26,
xtick style={color=black},
y grid style={white!69.0196078431373!black},
ylabel={BER},
ymajorgrids,
ymin=0.000214571953855644, ymax=0.0338105777382641,
ymode=log,
ytick style={color=black}
]
\addplot [semithick, color0, mark=+, mark size=3, mark options={solid}]
table {%
10 0.0242467953264713
12.5 0.0110496794804931
15 0.00532959401607513
17.5 0.00306666668038815
20 0.00180242673793275
22.5 0.0013253205235944
25 0.00101121795014478
};
\addplot [semithick, color1, mark=x, mark size=3, mark options={solid}]
table {%
10 0.0248493591696024
12.5 0.0113237178884447
15 0.00565117519969742
17.5 0.00348012819886208
20 0.00210485347945775
22.5 0.00157339741064029
25 0.00131458330112652
};
\addplot [semithick, color2, mark=asterisk, mark size=3, mark options={solid}]
table {%
10 0.0268637819588184
12.5 0.0128918271511793
15 0.00658119671667616
17.5 0.00401378201786429
20 0.00243475278174239
22.5 0.00198326209404816
25 0.00161121793207712
};
\addplot [semithick, color3, mark=square, mark size=3, mark options={solid}]
table {%
10 0.0185032039880753
12.5 0.00945833325386047
15 0.00454927887767553
17.5 0.00290643144398928
20 0.00159893196541816
22.5 0.00103966344613582
25 0.000605311302933842
};
\end{axis}

\end{tikzpicture}
			\vspace{-8pt}
	  		\caption{Coded BER with $\gpeak=9$~\si{dB}, $R=0$}
	  		\label{fig:ber_1}
		\end{subfigure}%
		\hfill
		\begin{subfigure}[b]{0.45\textwidth}
\begin{tikzpicture}

\definecolor{color0}{rgb}{0.172549019607843,0.627450980392157,0.172549019607843}
\definecolor{color1}{rgb}{1,0.498039215686275,0.0549019607843137}
\definecolor{color2}{rgb}{0.83921568627451,0.152941176470588,0.156862745098039}
\definecolor{color3}{rgb}{0.12156862745098,0.466666666666667,0.705882352941177}

\pgfplotsset{
    width=.88\textwidth,
    height=0.8\textwidth
}

\begin{axis}[
legend cell align={left},
legend style={
  fill opacity=0.8,
  draw opacity=1,
  text opacity=1,
  at={(0.03,0.97)},
  anchor=north west,
  draw=white!80!black
},
tick align=outside,
tick pos=left,
x grid style={white!69.0196078431373!black},
xlabel={SNR},
xmajorgrids,
xmin=9, xmax=31,
xtick style={color=black},
y grid style={white!69.0196078431373!black},
ylabel={Goodput [\si{\bit}]},
ymajorgrids,
ymin=1.80203300261452, ymax=2.00720599866869,
ytick style={color=black}
]
\addplot [semithick, color0, mark=+, mark size=3, mark options={solid}]
table {%
10 1.89266666299105
12.5 1.9382500013113
15 1.97181089781225
17.5 1.98249519238621
20 1.99067521343629
22.5 1.99215170954913
25 1.99550480760324
27.5 1.99692548075109
30 1.9978799533935
};
\addplot [semithick, color1, mark=x, mark size=3, mark options={solid}]
table {%
10 1.87918910384178
12.5 1.93209294646978
15 1.97097115404904
17.5 1.97987179485708
20 1.98986858967692
22.5 1.99137606840829
25 1.99476335469323
27.5 1.99613581730227
30 1.99694900933272
};
\addplot [semithick, color2, mark=asterisk, mark size=3, mark options={solid}]
table {%
10 1.86692307889462
12.5 1.92449038147926
15 1.96605448797345
17.5 1.97667467994243
20 1.9875128204003
22.5 1.98948717983067
25 1.99384989317817
27.5 1.99557972754701
30 1.9968234265583
};
\addplot [semithick, color3, mark=square, mark size=3, mark options={solid}]
table {%
10 1.81135904788971
12.5 1.87030231952667
15 1.89836847782135
17.5 1.90702819824219
20 1.91641187667847
22.5 1.9213844537735
25 1.92510640621185
27.5 1.92701888084412
30 1.92815291881561
};
\end{axis}

\end{tikzpicture}
			\vspace{-8pt}
			\caption{Goodput with $\gpeak=9$~\si{dB}, $R=0$}
			\label{fig:goodput_1}
		\end{subfigure}

		\vspace{10pt}

		\begin{subfigure}[b]{0.45\textwidth}
\begin{tikzpicture}

\definecolor{color0}{rgb}{0.172549019607843,0.627450980392157,0.172549019607843}
\definecolor{color1}{rgb}{1,0.498039215686275,0.0549019607843137}
\definecolor{color2}{rgb}{0.83921568627451,0.152941176470588,0.156862745098039}
\definecolor{color3}{rgb}{0.12156862745098,0.466666666666667,0.705882352941177}

\pgfplotsset{
    width=.88\textwidth,
    height=0.8\textwidth
}

\begin{axis}[
legend cell align={left},
legend style={fill opacity=0.8, draw opacity=1, text opacity=1, draw=white!80!black},
log basis y={10},
tick align=outside,
tick pos=left,
x grid style={white!69.0196078431373!black},
xlabel={$\frac{E_b}{\sigma^2}$},
xmajorgrids,
xtick={10, 12.5, 15, ..., 25},
xmin=9, xmax=26,
xtick style={color=black},
y grid style={white!69.0196078431373!black},
ylabel={BER},
ymajorgrids,
ymin=0.000221946495795994, ymax=0.028763574430894,
ymode=log,
ytick style={color=black}
]
\addplot [semithick, color0, mark=+, mark size=3, mark options={solid}]
table {%
10 0.0166458338499069
12.5 0.0101586541905999
15 0.00553060889393091
17.5 0.00324951919396408
20 0.00208899569871525
22.5 0.00159217033662966
25 0.0012090099850566
};
\addplot [semithick, color1, mark=x, mark size=3, mark options={solid}]
table {%
10 0.0164567306637764
12.5 0.0101009615510702
15 0.00581570521555841
17.5 0.00352123400615528
20 0.00234412391738345
22.5 0.00188438645896635
25 0.00140448717391118
};
\addplot [semithick, color2, mark=asterisk, mark size=3, mark options={solid}]
table {%
10 0.0230576917529106
12.5 0.0122692298442125
15 0.00761778855323792
17.5 0.00427003193181008
20 0.00293776709198331
22.5 0.00247802197866674
25 0.00180573363064064
};
\addplot [semithick, color3, mark=square, mark size=3, mark options={solid}]
table {%
10 0.017368588924408
12.5 0.009767628274858
15 0.00500881392508745
17.5 0.00298263877630234
20 0.00185845349915326
22.5 0.00114800815936178
25 0.000636587734334171
};
\end{axis}

\end{tikzpicture}
			\vspace{-8pt}
			\caption{Coded BER with $\gpeak=7$~\si{dB}, $R=4$}
			\label{fig:ber_2}
		\end{subfigure}%
		\hfill
	  	\begin{subfigure}[b]{0.45\textwidth}
\begin{tikzpicture}

\definecolor{color0}{rgb}{0.172549019607843,0.627450980392157,0.172549019607843}
\definecolor{color1}{rgb}{1,0.498039215686275,0.0549019607843137}
\definecolor{color2}{rgb}{0.83921568627451,0.152941176470588,0.156862745098039}
\definecolor{color3}{rgb}{0.12156862745098,0.466666666666667,0.705882352941177}

\pgfplotsset{
    width=.88\textwidth,
    height=0.8\textwidth
}

\begin{axis}[
legend cell align={left},
legend style={
  fill opacity=0.8,
  draw opacity=1,
  text opacity=1,
  at={(0.97,0.03)},
  anchor=south east,
  draw=white!80!black
},
tick align=outside,
tick pos=left,
x grid style={white!69.0196078431373!black},
xlabel={SNR},
xmajorgrids,
xmin=9, xmax=31,
xtick style={color=black},
y grid style={white!69.0196078431373!black},
ylabel={Goodput [\si{\bit}]},
ymajorgrids,
ymin=1.70317240780278, ymax=2.01090700686618,
ytick style={color=black}
]
\addplot [semithick, color0, mark=+, mark size=3, mark options={solid}]
table {%
10 1.86943909525871
12.5 1.94126923009753
15 1.95630769059062
17.5 1.97447435733676
20 1.98432371811941
22.5 1.98979807707171
25 1.99287019227631
27.5 1.99581089743879
30 1.99649759609019
};
\addplot [semithick, color1, mark=x, mark size=3, mark options={solid}]
table {%
10 1.85358654499054
12.5 1.92886858969927
15 1.95584615319967
17.5 1.97449679476023
20 1.98471474321559
22.5 1.99001923048248
25 1.99363862187602
27.5 1.99609668808989
30 1.99691907054512
};
\addplot [semithick, color2, mark=asterisk, mark size=3, mark options={solid}]
table {%
10 1.80550000071526
12.5 1.90964423120022
15 1.93933974206448
17.5 1.9657948705256
20 1.97988621797413
22.5 1.98665491460512
25 1.99149919860065
27.5 1.99446474364959
30 1.99571674680919
};
\addplot [semithick, color3, mark=square, mark size=3, mark options={solid}]
table {%
10 1.71716034412384
12.5 1.76436686515808
15 1.79517424106598
17.5 1.80874693393707
20 1.81268966197968
22.5 1.81943321228027
25 1.82268881797791
27.5 1.82428359985352
30 1.82536792755127
};
\end{axis}

\end{tikzpicture}
			\vspace{-8pt}
			\caption{Goodput with $\gpeak=7$~\si{dB}, $R=4$}
	  		\label{fig:goodput_2}
		\end{subfigure}%

		\vspace{10pt}

		\begin{subfigure}[b]{0.45\textwidth}
\begin{tikzpicture}

\definecolor{color0}{rgb}{0.172549019607843,0.627450980392157,0.172549019607843}
\definecolor{color1}{rgb}{1,0.498039215686275,0.0549019607843137}
\definecolor{color2}{rgb}{0.83921568627451,0.152941176470588,0.156862745098039}
\definecolor{color3}{rgb}{0.12156862745098,0.466666666666667,0.705882352941177}

\pgfplotsset{
    width=.88\textwidth,
    height=0.8\textwidth
}

\begin{axis}[
legend cell align={left},
legend style={fill opacity=0.8, draw opacity=1, text opacity=1, draw=white!80!black},
log basis y={10},
tick align=outside,
tick pos=left,
x grid style={white!69.0196078431373!black},
xlabel={$\frac{E_b}{\sigma^2}$},
xmajorgrids,
xtick={10, 12.5, 15, ..., 25},
xmin=9, xmax=26,
xtick style={color=black},
y grid style={white!69.0196078431373!black},
ylabel={BER},
ymajorgrids,
ymin=0.000228713345792401, ymax=0.0360773952572388,
ymode=log,
ytick style={color=black}
]
\addplot [semithick, color0, mark=+, mark size=3, mark options={solid}]
table {%
10 0.0183685906231403
12.5 0.0124326932281256
15 0.00780689087696373
17.5 0.0044959937222302
20 0.00292040597802649
22.5 0.00221367530028025
25 0.00194262822798143
};
\addplot [semithick, color1, mark=x, mark size=3, mark options={solid}]
table {%
10 0.0230945521295071
12.5 0.015955128595233
15 0.0100328526459634
17.5 0.00542868603952229
20 0.00346923075287292
22.5 0.00278130341495077
25 0.00221933758712063
};
\addplot [semithick, color2, mark=asterisk, mark size=3, mark options={solid}]
table {%
10 0.0286634620279074
12.5 0.0214022441804409
15 0.0145592945627868
17.5 0.00985016021877527
20 0.00768803401539723
22.5 0.00666987181951602
25 0.00596955139189959
};
\addplot [semithick, color3, mark=square, mark size=3, mark options={solid}]
table {%
10 0.0191025640815496
12.5 0.012591346167028
15 0.00727243581786752
17.5 0.00368397426791489
20 0.00209655473008752
22.5 0.00134842412080616
25 0.000792200909927487
};
\end{axis}

\end{tikzpicture}
			\vspace{-8pt}
			\caption{Coded BER with $\gpeak=5$~\si{dB}, $R=16$}
			\label{fig:ber_3}
		\end{subfigure}
		\hfill
		\begin{subfigure}[b]{0.45\textwidth}
\begin{tikzpicture}

\definecolor{color0}{rgb}{0.172549019607843,0.627450980392157,0.172549019607843}
\definecolor{color1}{rgb}{1,0.498039215686275,0.0549019607843137}
\definecolor{color2}{rgb}{0.83921568627451,0.152941176470588,0.156862745098039}
\definecolor{color3}{rgb}{0.12156862745098,0.466666666666667,0.705882352941177}

\pgfplotsset{
    width=.88\textwidth,
    height=0.8\textwidth
}

\begin{axis}[
legend cell align={left},
legend style={
  fill opacity=0.8,
  draw opacity=1,
  text opacity=1,
  at={(0.91,0.5)},
  anchor=east,
  draw=white!80!black
},
tick align=outside,
tick pos=left,
x grid style={white!69.0196078431373!black},
xlabel={SNR},
xmajorgrids,
xmin=9, xmax=31,
xtick style={color=black},
y grid style={white!69.0196078431373!black},
ylabel={Goodput [\si{\bit}]},
ymajorgrids,
ymin=1.38662141720221, ymax=2.02258589511039,
ytick style={color=black}
]
\addplot [semithick, color0, mark=+, mark size=3, mark options={solid}]
table {%
10 1.82704167068005
12.5 1.90628525614738
15 1.94774359092116
17.5 1.97000000067055
20 1.98129006409645
22.5 1.99007371789776
25 1.99266025656834
27.5 1.99367841884183
30 1.9925016022753
};
\addplot [semithick, color1, mark=x, mark size=3, mark options={solid}]
table {%
10 1.81219550848007
12.5 1.89268909811974
15 1.93866666615009
17.5 1.95922756567597
20 1.97620192383975
22.5 1.98803525650874
25 1.99264423083514
27.5 1.99354807691028
30 1.99304326926358
};
\addplot [semithick, color2, mark=asterisk, mark size=3, mark options={solid}]
table {%
10 1.78941987454891
12.5 1.87328526377678
15 1.92434615641832
17.5 1.95166666805744
20 1.96713782036304
22.5 1.97843108978122
25 1.98350320477039
27.5 1.98550641008963
30 1.98374679498374
};
\addplot [semithick, color3, mark=square, mark size=3, mark options={solid}]
table {%
10 1.41552889347076
12.5 1.46134424209595
15 1.48576760292053
17.5 1.50154948234558
20 1.50540804862976
22.5 1.51302576065063
25 1.51444876194
27.5 1.51605880260468
30 1.51698064804077
};
\end{axis}

\end{tikzpicture}
			\vspace{-8pt}
			\caption{Goodput with $\gpeak=5$~\si{dB}, $R=16$}
			\label{fig:goodput_3}
		\end{subfigure}%
	
	\caption{BER and goodput achieved by the different schemes.}
	\label{fig:ber_goodput}
	\end{figure}
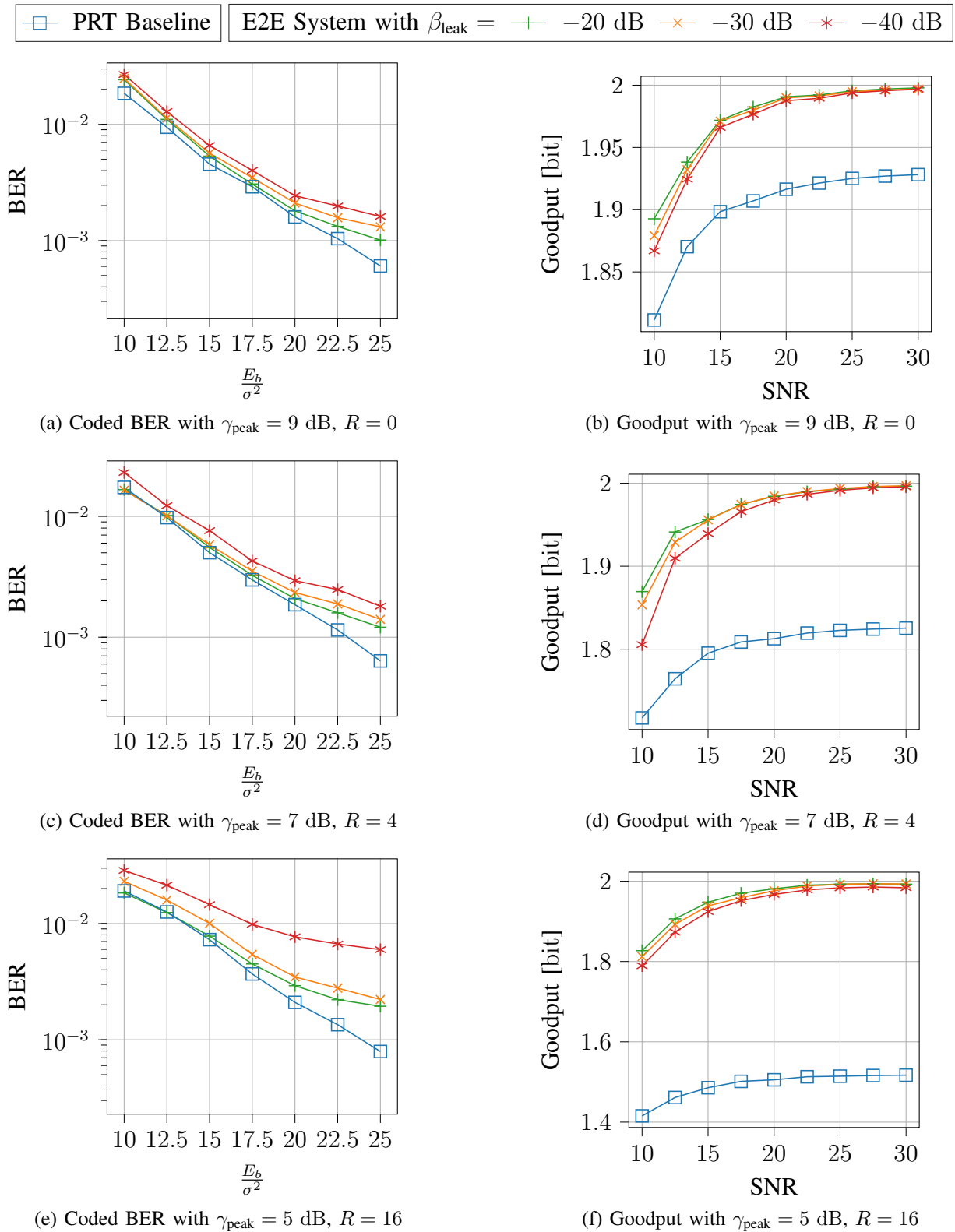

The coded \glspl{BER} of the baseline and of the three E2E systems are shown in the left column of Fig.~\ref{fig:ber_goodput}.
As the baseline transmits pilots and reduction signals in addition to data signals, the energy per transmitted bit is higher than that of the of E2E systems.
To reflect this characteristic, we define the energy-per-bit-to-noise-spectral-density ratio as
\begin{align}
	\frac{E_b}{\sigma^2} = \frac{\EE_{x_{m,n}} \left[ |x_{m,n}|^2 \right]}{\rho K \sigma^2} = \frac{1}{\rho K \sigma^2}
\end{align}
where $\rho$ is the ratio of \glspl{RE} carrying data signals over the total number of \glspl{RE} in the \gls{RG}.
Fig.~\ref{fig:ber_1},~\ref{fig:ber_2}, and~\ref{fig:ber_3} correspond to systems achieving \glspl{PAPR} of approximately $8.5$, $6.8$, and $5.2~\si{dB}$, respectively.
Such \glspl{PAPR} were obtained using $R=\{0, 4, 16\}$ \gls{PRT} for the baseline, and $\gpeak = \{9, 7, 5\}~\si{dB}$ for the trained systems.
Note that evaluation results corresponding to systems trained with $\bleak = \{-30, -40\}$~\si{dB} are provided for completeness, but a fair comparison is only possible between the baseline and the system trained with $\bleak = -20$~\si{dB} as this value corresponds to the baseline \gls{ACLR}.
Overall, one can see that the baseline consistently achieves slightly lower \gls{BER} than the E2E systems.
However, the baseline also transmits fewer bits per \gls{RG}, due to some \glspl{RE} being used to transmit pilots or reduction signals.

To understand the benefits provided by the E2E approach, the second column of Fig.~\ref{fig:ber_goodput} presents the \emph{goodputs} achieved by each compared system.
The goodput is defined as the average number of information bits that have been successfully received in a RE, i.e., 
\begin{align}
	\text{Goodput} = r \rho K (1-\text{BER}),
\end{align}
and is plotted with respect to the SNR as it already accounts for the different number of \glspl{RE} transmitting information bits through the parameter $\rho$.
Evaluation results show that the goodputs achieved by all trained systems, including those trained for lower \glspl{ACLR}, are significantly higher than the ones of the baseline. 
Indeed, at an SNR of $30~\si{dB}$, all E2E systems are able to successfully transmit close to two bits per RE, while the baseline saturates at $1.93$, $1.82$, and $1.53$ bits for \glspl{PAPR} of $8.5$, $6.8$, and $5.2$ \si{dB}, respectively.
The BER improvements range from a $3\%$ increase at low SNR with $\text{PAPR}_{\epsilon} \approx 8.5~\si{dB}$ to a $30\%$ increase at high SNR with $\text{PAPR}_{\epsilon} \approx 5.2~\si{dB}$, indicating that the E2E approach is particularly effective when the \gls{PAPR} reduction is high.
These gains are jointly enabled by the pilotless nature of the E2E transmission, the effective \gls{ACLR} reduction scheme learned through the proposed optimization procedure, and the fact that every \glspl{RE} can be used to transmit data.

\subsection{Insights on the learned \gls{ACLR} and \gls{PAPR} reduction techniques}

\subsubsection{Interpreting the \gls{ACLR} minimization process}

\begin{figure} 
	\centering 
	  \begin{subfigure}[b]{0.20\linewidth}
\begin{tikzpicture}

\begin{axis}[
width=5cm,height=5cm,
tick align=outside,
tick pos=left,
x grid style={white!69.0196078431373!black},
xmin=-0.5, xmax=74.5,
xtick style={color=black},
xtick={7,22,37,52,67},
xticklabels={-30,-15,0,15,30},
y dir=reverse,
y grid style={white!69.0196078431373!black},
ymin=-0.5, ymax=74.5,
ytick style={color=black},
ytick={7,22,37,52,67},
yticklabels={-30,-15,0,15,30}
]
\addplot graphics [includegraphics cmd=\pgfimage,xmin=-0.5, xmax=74.5, ymin=74.5, ymax=-0.5] {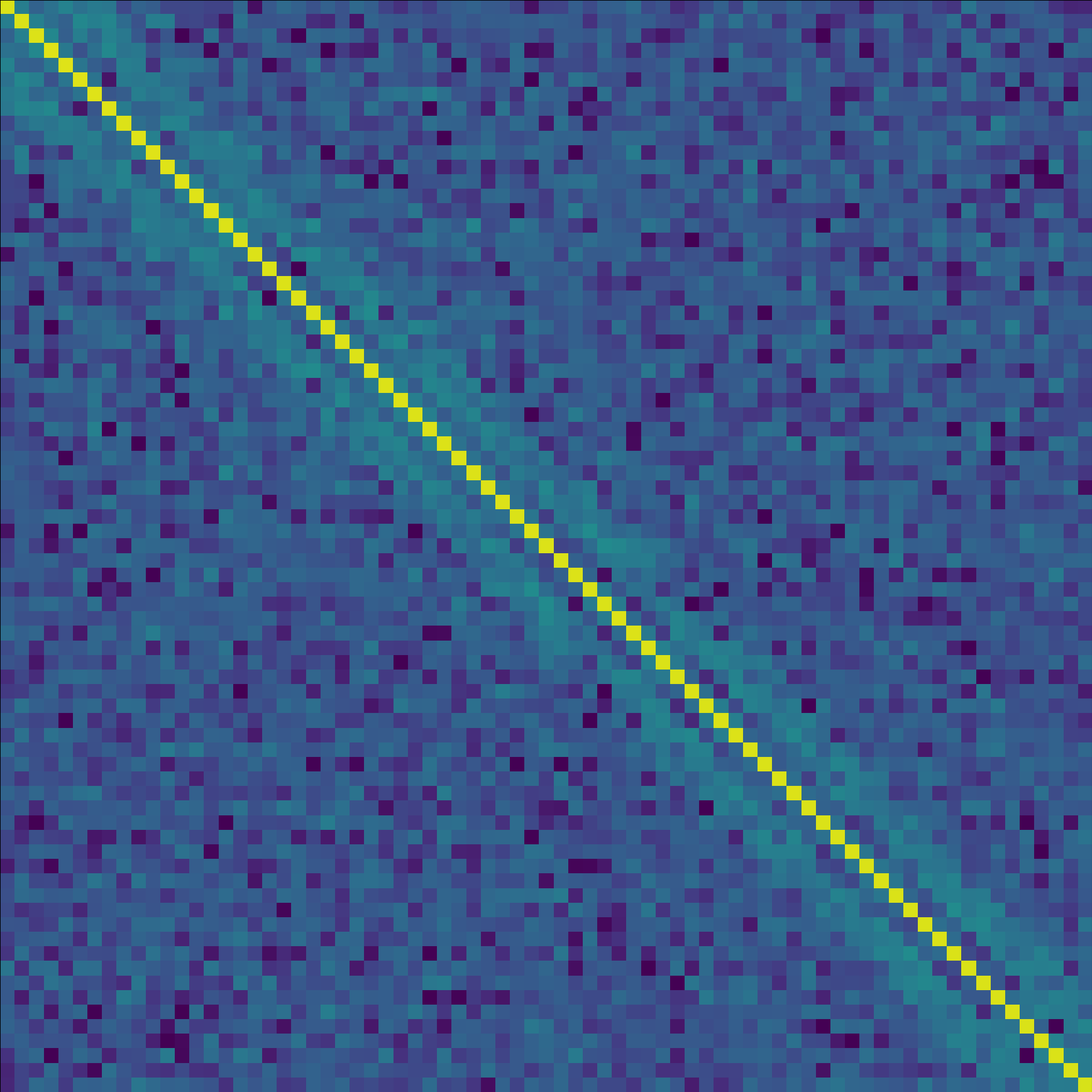};
\end{axis}

\end{tikzpicture}
		\vspace{-25pt}
		\caption{$\bleak=-20~\si{dB}$} \label{fig:M1}  
	  \end{subfigure}
	  \hspace{30pt}
	\begin{subfigure}[b]{0.20\linewidth}
\begin{tikzpicture}

\begin{axis}[
width=5cm,height=5cm,
tick align=outside,
tick pos=left,
x grid style={white!69.0196078431373!black},
xmin=-0.5, xmax=74.5,
xtick style={color=black},
xtick={7,22,37,52,67},
xticklabels={-30,-15,0,15,30},
y dir=reverse,
y grid style={white!69.0196078431373!black},
ymin=-0.5, ymax=74.5,
ytick style={color=black},
ytick={7,22,37,52,67},
yticklabels={-30,-15,0,15,30}
]
\addplot graphics [includegraphics cmd=\pgfimage,xmin=-0.5, xmax=74.5, ymin=74.5, ymax=-0.5] {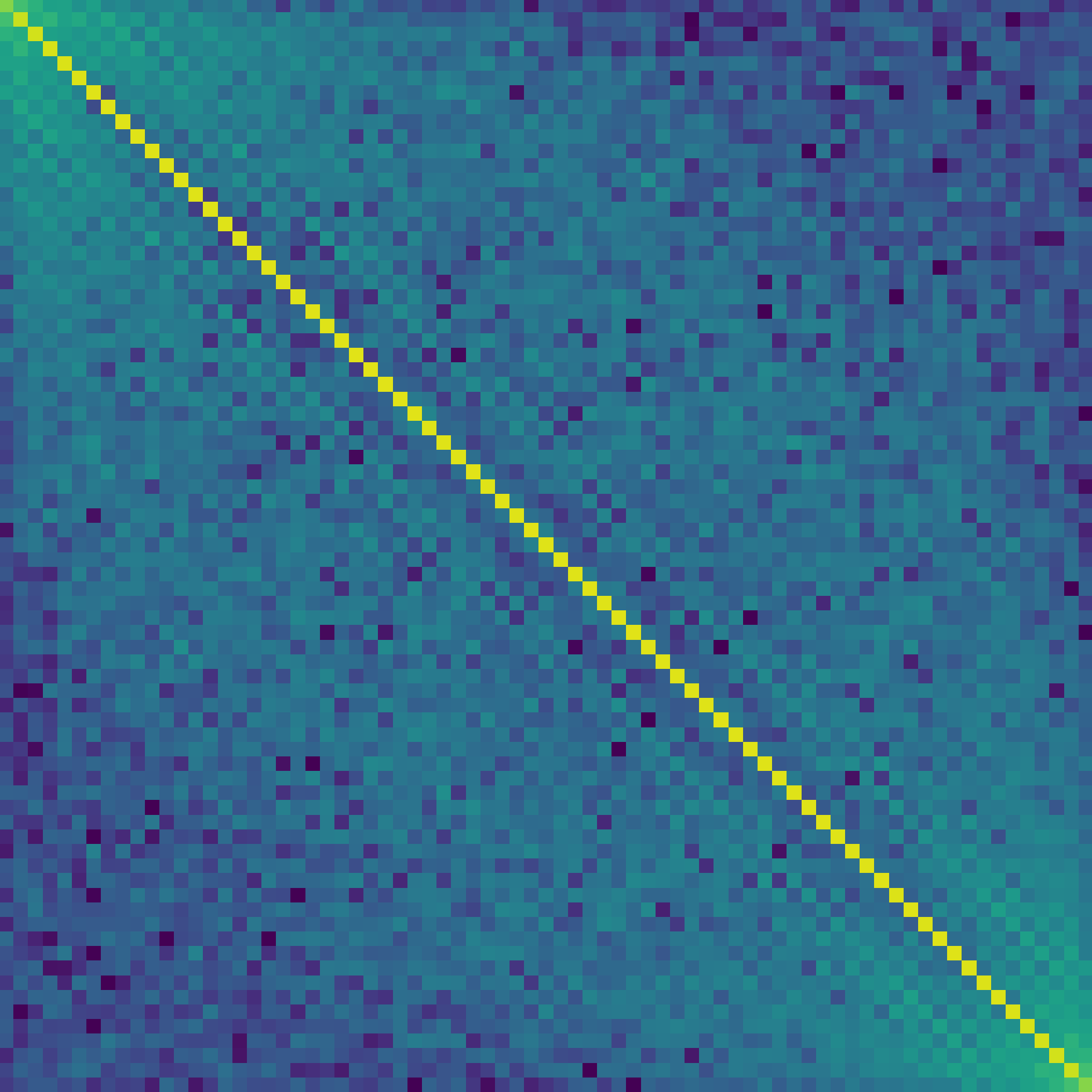};
\end{axis}

\end{tikzpicture}
	  \vspace{-25pt}
	\caption{$\bleak=-30~\si{dB}$} \label{fig:M2}  
	\end{subfigure}
	\hspace{30pt}
	\begin{subfigure}[b]{0.20\linewidth}
\begin{tikzpicture}

\begin{axis}[
width=5cm,height=5cm,
tick align=outside,
tick pos=left,
x grid style={white!69.0196078431373!black},
xmin=-0.5, xmax=74.5,
xtick style={color=black},
xtick={7,22,37,52,67},
xticklabels={-30,-15,0,15,30},
y dir=reverse,
y grid style={white!69.0196078431373!black},
ymin=-0.5, ymax=74.5,
ytick style={color=black},
ytick={7,22,37,52,67},
yticklabels={-30,-15,0,15,30}
]
\addplot graphics [includegraphics cmd=\pgfimage,xmin=-0.5, xmax=74.5, ymin=74.5, ymax=-0.5] {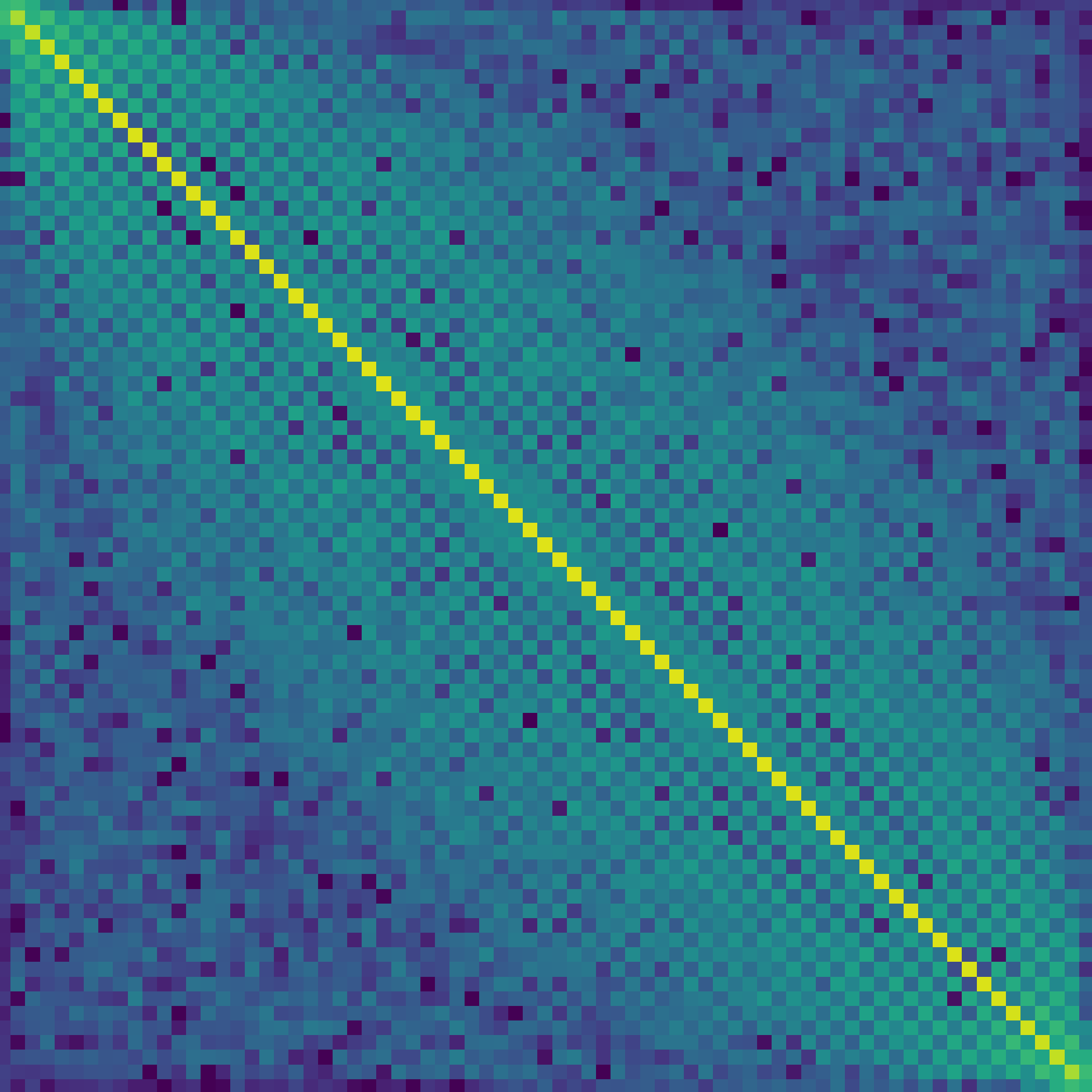};
\end{axis}

\end{tikzpicture}
		\vspace{-25pt}
	  \caption{$\bleak=-40~\si{dB}$} \label{fig:M3}  
	  \end{subfigure}
	  \hspace{35pt}
	  \begin{subfigure}[b]{0.15\linewidth}
\begin{tikzpicture}

\begin{axis}[
width=2cm,height=5cm,
axis y line=right,
tick align=outside,
xmin=-30, xmax=1.5,
xtick pos=left,
xtick=\empty,
y grid style={white!69.0196078431373!black},
ymin=-30, ymax=1.5,
ytick pos=right,
ylabel={Correlation [\si{\dB}]},
ytick style={color=black},
yticklabel style={anchor=west}
]
\addplot graphics [includegraphics cmd=\pgfimage,xmin=-30, xmax=1.5, ymin=-30, ymax=1.5] {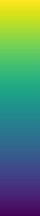};
\end{axis}

\end{tikzpicture}
		\vspace{32pt}
	  \end{subfigure}
	\caption{Covariance matrices $\EE_m \left[\xv_m \xv_m \htp \right]$ for systems trained with different target \glspl{ACLR}.}
	\label{fig:cov_subcarriers}
\end{figure}

In order to get insight into the processing done by the neural transmitter, we first study the \gls{ACLR} reduction technique learned by the E2E systems.
Three covariance matrices $\EE_{\xv_m} \left[\xv_m \xv_m \htp \right]$ are shown in Fig.~\ref{fig:cov_subcarriers} for systems trained with $\gpeak = 9$.
The first covariance matrix is from a system trained with an \gls{ACLR} target of $\bleak=-20~\si{dB}$, while the second and third matrices were respectively obtained with $\bleak=-30~\si{dB}$ and $\bleak=-40~\si{dB}$.
It can be observed that the correlation between the elements close to the diagonal increase inversely to the \gls{ACLR} target. 
Moreover, the correlation increases at subcarriers located near the edges of the spectrum, indicating that a subcarrier-dependent filtering is learned.
Fig.~\ref{fig:energy_subcarriers} depicts the distribution of the energy across the subcarriers. 
On the one hand, the system trained with a lax \gls{ACLR} constraint ($\bleak=-20~\si{dB}$) equally distributes the available energy on all subcarriers, which can be shown to maximize the information rate of the transmission.
On the other hand, the systems trained with lower \gls{ACLR} targets learn to reduce the energy of the subcarriers located at the \gls{RG} edges, also contributing the out-of-band emissions reduction. 
The joint effect of the filtering and of the uneven energy distribution across subcarriers is visible in Fig.~\ref{fig:psd}, where the \glspl{PSD} of the compared systems are shown.
One can see that the system trained with $\bleak=-20~\si{dB}$ and the baseline have similar \glspl{PSD}, while the \glspl{PSD} of the systems trained with lower \gls{ACLR} targets present significantly lower out-of-band emissions.
It therefore appears that the improved \gls{ACLR} allowed by the E2E system is the result of both reducing the power of the subcarriers located near the \gls{RG} edges and the introduction of correlations between the transmitted symbols.

\begin{figure}
	\centering
	\begin{minipage}{.45\textwidth}
	  \centering
\begin{tikzpicture}

\definecolor{color0}{rgb}{1,0.498039215686275,0.0549019607843137}
\definecolor{color1}{rgb}{0.172549019607843,0.627450980392157,0.172549019607843}
\definecolor{color2}{rgb}{0.83921568627451,0.152941176470588,0.156862745098039}

\pgfplotsset{
    width=.88\textwidth,
    height=0.8\textwidth
}

\begin{axis}[
legend cell align={left},
legend style={
  fill opacity=0.8,
  draw opacity=1,
  text opacity=1,
  at={(0.5,0.02)},
  anchor=south,
  draw=white!80!black
},
tick align=outside,
tick pos=left,
x grid style={white!69.0196078431373!black},
xlabel={Subcarrier index},
xtick = {-30, -15, 0, 15, 30},
xmajorgrids,
xmin=-40.7, xmax=40.7,
xtick style={color=black},
y grid style={white!69.0196078431373!black},
ylabel={$E_n \left[|\xv_n|^2 \right]$},
ymajorgrids,
ymin=0.0576631758362055, ymax=1.12922138385475,
ytick style={color=black}
]
\addplot [semithick, color1]
table {%
-37 1.03111910820007
-36 1.03385746479034
-35 1.03923320770264
-34 1.05720007419586
-33 1.03550052642822
-32 1.03485810756683
-31 1.02800130844116
-30 1.04845309257507
-29 1.02976274490356
-28 1.01962292194366
-27 1.0232492685318
-26 1.01172542572021
-25 1.02883803844452
-24 1.02799785137177
-23 1.01899302005768
-22 1.02556192874908
-21 1.00689446926117
-20 1.01758122444153
-19 1.03649234771729
-18 1.00674319267273
-17 1.01164877414703
-16 0.988463521003723
-15 1.00538897514343
-14 1.00149941444397
-13 1.01179563999176
-12 0.997092366218567
-11 0.986716389656067
-10 0.996045649051666
-9 1.00028884410858
-8 0.99087780714035
-7 0.998530089855194
-6 0.993778645992279
-5 0.990337729454041
-4 0.993604958057404
-3 0.984837710857391
-2 0.989624381065369
-1 0.997469842433929
0 0.996970355510712
1 0.978860855102539
2 1.006711602211
3 0.981540203094482
4 0.995557904243469
5 0.980491936206818
6 0.985247373580933
7 0.995787739753723
8 0.986366927623749
9 0.977036476135254
10 0.983359217643738
11 0.980548322200775
12 0.985097825527191
13 0.99271833896637
14 0.978238046169281
15 0.97946685552597
16 1.0110969543457
17 0.981496870517731
18 0.994325816631317
19 0.98721307516098
20 0.968293905258179
21 0.975181102752686
22 0.982566893100739
23 0.996837615966797
24 0.970447838306427
25 0.982376873493195
26 0.956460475921631
27 0.982670068740845
28 0.971469819545746
29 0.987303018569946
30 0.989450216293335
31 0.995298802852631
32 0.98803448677063
33 1.00204539299011
34 1.00887608528137
35 1.01165330410004
36 0.996788799762726
37 0.97643107175827
};
\addlegendentry{$\bleak = -20$~\si{dB}}
\addplot [semithick, color0]
table {%
-37 0.394234329462051
-36 0.849884510040283
-35 0.92091315984726
-34 0.979568660259247
-33 0.991887509822845
-32 0.994583189487457
-31 1.00772309303284
-30 1.02929127216339
-29 1.03279507160187
-28 1.01638829708099
-27 1.01908099651337
-26 1.01274573802948
-25 1.02495062351227
-24 1.03006303310394
-23 1.02270567417145
-22 1.021613240242
-21 1.03020024299622
-20 1.03042006492615
-19 1.05113446712494
-18 1.02504181861877
-17 1.03956985473633
-16 1.00732982158661
-15 1.02720677852631
-14 1.04590737819672
-13 1.04506099224091
-12 1.02441620826721
-11 1.02510559558868
-10 1.03288722038269
-9 1.03358221054077
-8 1.02628362178802
-7 1.02852845191956
-6 1.03796494007111
-5 1.03753614425659
-4 1.04288172721863
-3 1.03350591659546
-2 1.03495895862579
-1 1.05018126964569
0 1.04681944847107
1 1.02783560752869
2 1.05139660835266
3 1.03249883651733
4 1.04407346248627
5 1.0327330827713
6 1.02391576766968
7 1.03990137577057
8 1.02911198139191
9 1.03274619579315
10 1.03038704395294
11 1.02464592456818
12 1.0388218164444
13 1.03748691082001
14 1.01784753799438
15 1.02575790882111
16 1.05120384693146
17 1.02341520786285
18 1.03270137310028
19 1.03154277801514
20 1.00719702243805
21 1.01436567306519
22 1.01709127426147
23 1.02963781356812
24 1.01677489280701
25 1.00660479068756
26 0.996297240257263
27 1.01689457893372
28 1.00580310821533
29 1.02261757850647
30 1.00837647914886
31 1.01957356929779
32 0.993280827999115
33 0.979392886161804
34 0.96235316991806
35 0.925103783607483
36 0.855167508125305
37 0.39049044251442
};
\addlegendentry{$\bleak = -30$~\si{dB}}
\addplot [semithick, color2]
table {%
-37 0.127902194857597
-36 0.688325762748718
-35 0.871064901351929
-34 0.875893712043762
-33 0.978251099586487
-32 0.963384985923767
-31 1.0073789358139
-30 1.00741195678711
-29 1.04175639152527
-28 1.00433719158173
-27 1.04092276096344
-26 1.02410411834717
-25 1.04981791973114
-24 1.02981007099152
-23 1.04763305187225
-22 1.04229605197906
-21 1.04704940319061
-20 1.0460296869278
-19 1.06218063831329
-18 1.05800700187683
-17 1.06249558925629
-16 1.0452653169632
-15 1.05291306972504
-14 1.05265533924103
-13 1.05029153823853
-12 1.06049919128418
-11 1.05081748962402
-10 1.07054913043976
-9 1.0497944355011
-8 1.06423544883728
-7 1.06710660457611
-6 1.07795822620392
-5 1.07040357589722
-4 1.06327962875366
-3 1.05679059028625
-2 1.06900668144226
-1 1.08051419258118
0 1.05856394767761
1 1.07525455951691
2 1.06486821174622
3 1.04599821567535
4 1.05602157115936
5 1.07804799079895
6 1.07512998580933
7 1.03510010242462
8 1.07397902011871
9 1.04767179489136
10 1.05928432941437
11 1.06938004493713
12 1.05251741409302
13 1.06591737270355
14 1.03210031986237
15 1.05259943008423
16 1.06347107887268
17 1.01992678642273
18 1.04817008972168
19 1.03191292285919
20 1.03644680976868
21 1.0608081817627
22 1.02749228477478
23 1.06456875801086
24 1.04929208755493
25 1.0314964056015
26 1.02537178993225
27 1.05142021179199
28 1.01878750324249
29 1.01814031600952
30 0.996125042438507
31 1.00453579425812
32 0.984402120113373
33 0.986095011234283
34 0.889274179935455
35 0.832385182380676
36 0.652933478355408
37 0.106370367109776
};
\addlegendentry{$\bleak = -40$~\si{dB}}
\end{axis}

\end{tikzpicture}
	  \captionof{figure}{Mean energy of the symbols transmitted on each subcarrier.}
		\vspace{-20pt}
	  \label{fig:energy_subcarriers}
	\end{minipage}%
	\hfill
	\begin{minipage}{.45\textwidth}
	  \centering
	  \input{eval/PSD_smooth.tex}
	  \captionof{figure}{PSD of four systems having no \gls{PAPR} constraint.}
		\vspace{-20pt}	
	  \label{fig:psd}
	\end{minipage}
\end{figure}

\subsubsection{Interpreting the \gls{PAPR} minimization process}
\begin{figure}[t]
	\centering
	\includegraphics[width=1\textwidth]{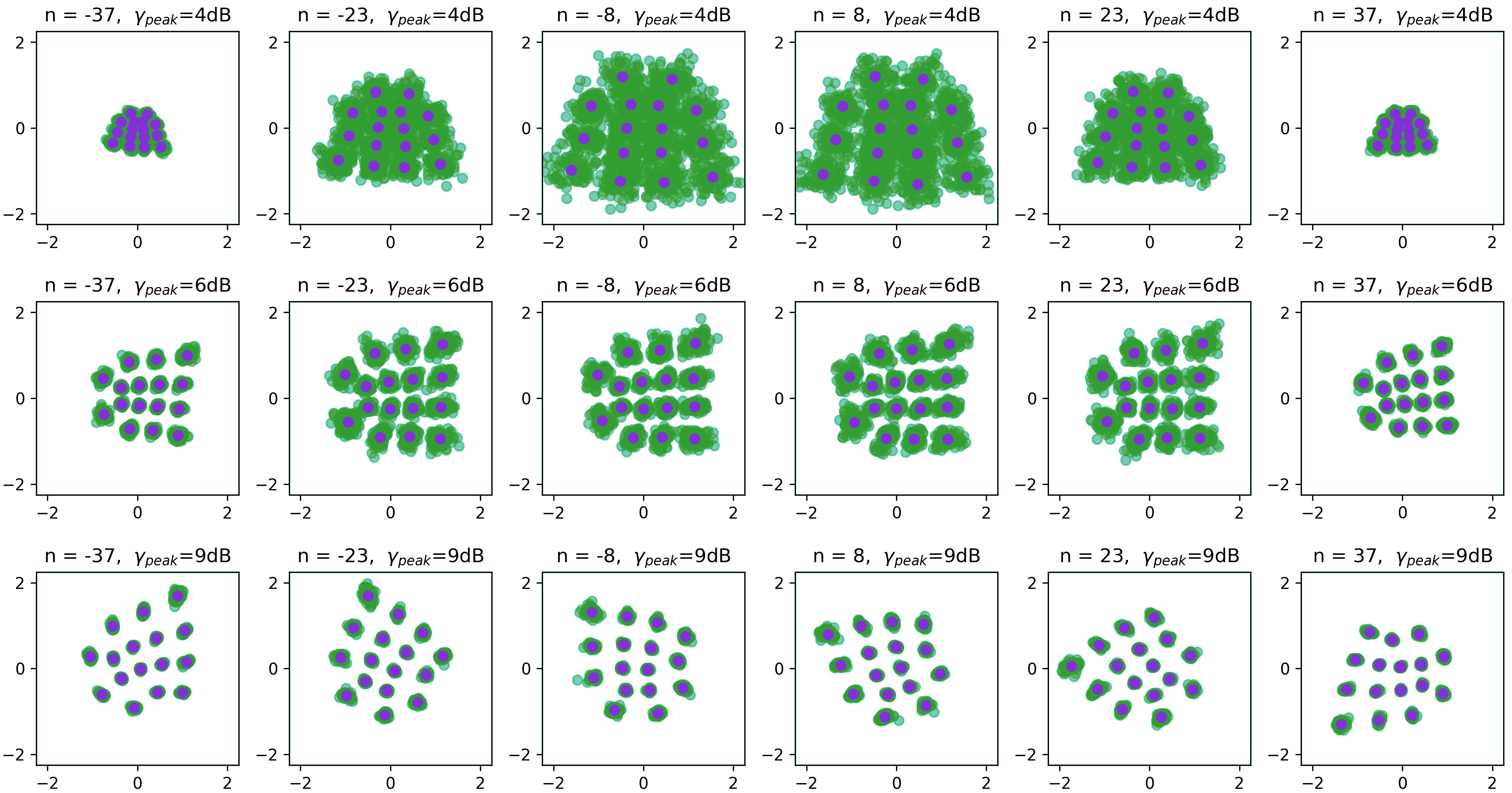}
	\caption{Transmitted signals on six subcarriers for E2E systems trained with different \gls{PAPR} targets but a lax \gls{ACLR} constraint. The purple dots represent the center of each cluster.}
	\label{fig:clusters}
\end{figure}

To understand how the neural transmitter and the optimization procedure enable significant \gls{PAPR} reduction, it is insightful to look at the symbols transmitted for different \gls{PAPR} targets, as illustrated in Fig.~\ref{fig:clusters}.
The three rows represent E2E systems trained for \gls{PAPR} targets of $\gpeak\in\{4, 6, 9\}~\si{dB}$, and the six columns show the signals transmitted on the subcarriers $n\in\{-37, -23, -8, 8, 23, 37\}$.
All systems where trained with a lax \gls{ACLR} constraint ($\bleak=-20\si{dB}$).
This figure was obtained by sending 25 \glspl{RG}, and each green point represents one signal sent on the corresponding subcarrier. 
One can see that the signals are gathered into $2^K$ groups, that will be referred to as \emph{clusters} in the following.
It can be observed that for low \gls{PAPR} targets ($\gpeak=4\si{dB}$), the clusters at the subcarriers located at the center of the \gls{RG} exhibit more dispersion than the ones located near the \gls{RG} edges. 
Moreover, the average energy of the transmitted signals also appears higher on the central subcarriers.
On the contrary, for high \gls{PAPR} targets ($\gpeak=9\si{dB}$), the clusters seem equally dispersed and the energy is evenly spread across the subcarriers.
Finally, it can be noted that the positions of the clusters are not rotationally symmetrical, which should help the neural receiver in estimating and equalizing the channel.

\begin{figure}
	\centering
	\begin{minipage}{.45\textwidth}
	  \centering
\begin{tikzpicture}

\definecolor{color0}{rgb}{0.172549019607843,0.627450980392157,0.172549019607843}

\pgfplotsset{
    width=.88\textwidth,
    height=0.8\textwidth
}

\begin{axis}[
legend cell align={left},
legend style={
  fill opacity=0.8,
  draw opacity=1,
  text opacity=1,
  at={(0.98,0.02)},
  anchor=south east,
  draw=white!80!black
},
tick align=outside,
tick pos=left,
x grid style={white!69.0196078431373!black},
xlabel={Subcarrier index $n$},
xmajorgrids,
xtick = {-30, -15, 0, 15, 30},
xmin=-40.7, xmax=40.7,
xtick style={color=black},
y grid style={white!69.0196078431373!black},
ylabel={$\EE_k \left[ |\bar{\mathcal{C}}(n,k)|^2 \right]$},
ymajorgrids,
ymin=0.0892762571573257, ymax=1.65094530284405,
ytick style={color=black}
]
\addplot [semithick, color0, dash pattern=on 1pt off 1pt]
table {%
-37 0.160261213779449
-36 0.178962647914886
-35 0.204348176717758
-34 0.230534255504608
-33 0.265210866928101
-32 0.296909153461456
-31 0.337118536233902
-30 0.376962333917618
-29 0.422083824872971
-28 0.468639135360718
-27 0.525643825531006
-26 0.584878146648407
-25 0.637288212776184
-24 0.685698926448822
-23 0.750823616981506
-22 0.810682594776154
-21 0.872012197971344
-20 0.927743196487427
-19 0.989519357681274
-18 1.04258060455322
-17 1.10673856735229
-16 1.1493968963623
-15 1.185462474823
-14 1.22462666034698
-13 1.27899265289307
-12 1.30964183807373
-11 1.34447550773621
-10 1.38905835151672
-9 1.41250479221344
-8 1.44237041473389
-7 1.47729229927063
-6 1.49166369438171
-5 1.53108608722687
-4 1.53331661224365
-3 1.56028628349304
-2 1.55523979663849
-1 1.57996034622192
0 1.57294714450836
1 1.5439178943634
2 1.55541825294495
3 1.55014157295227
4 1.53359138965607
5 1.52383852005005
6 1.51304030418396
7 1.48090636730194
8 1.46889901161194
9 1.44764363765717
10 1.39895272254944
11 1.39651536941528
12 1.33268117904663
13 1.30107545852661
14 1.27028703689575
15 1.22734475135803
16 1.17434978485107
17 1.13555085659027
18 1.06350696086884
19 1.02035462856293
20 0.957301497459412
21 0.898664116859436
22 0.827974379062653
23 0.763263463973999
24 0.703479945659637
25 0.643402338027954
26 0.579641342163086
27 0.533875226974487
28 0.478126436471939
29 0.424997538328171
30 0.375556796789169
31 0.336856544017792
32 0.301724493503571
33 0.267655521631241
34 0.233090728521347
35 0.20718240737915
36 0.181130275130272
37 0.160344213247299
};
\addlegendentry{$\gpeak=4$}
\addplot [semithick, color0, dashed]
table {%
-37 0.710524439811707
-36 0.745990812778473
-35 0.774487495422363
-34 0.791763842105865
-33 0.818379938602448
-32 0.852114379405975
-31 0.878749489784241
-30 0.904673337936401
-29 0.940613150596619
-28 0.956783831119537
-27 0.99269038438797
-26 0.986374020576477
-25 1.00299847126007
-24 1.01407110691071
-23 1.03190755844116
-22 1.04131293296814
-21 1.0415552854538
-20 1.04279935359955
-19 1.04042935371399
-18 1.03951811790466
-17 1.03398036956787
-16 1.03911423683167
-15 1.02434182167053
-14 1.02074670791626
-13 1.02825307846069
-12 1.02809679508209
-11 1.02226865291595
-10 1.02563858032227
-9 1.03517019748688
-8 1.04234409332275
-7 1.04695785045624
-6 1.04857695102692
-5 1.05702042579651
-4 1.05142641067505
-3 1.05929148197174
-2 1.06771230697632
-1 1.07334017753601
0 1.0705099105835
1 1.06966495513916
2 1.05558276176453
3 1.05743813514709
4 1.05309534072876
5 1.0497499704361
6 1.03988826274872
7 1.04067420959473
8 1.03577184677124
9 1.02498316764832
10 1.01053285598755
11 1.00752139091492
12 1.01200258731842
13 1.00238311290741
14 1.0101752281189
15 1.00128769874573
16 1.00653457641602
17 1.00986409187317
18 1.02007591724396
19 1.02245712280273
20 1.03194296360016
21 1.03731513023376
22 1.02125322818756
23 1.01891994476318
24 1.008953332901
25 0.999530911445618
26 0.986116647720337
27 0.965908765792847
28 0.946252524852753
29 0.92317807674408
30 0.897688984870911
31 0.877076089382172
32 0.842434763908386
33 0.819824934005737
34 0.786785185337067
35 0.764299929141998
36 0.754411935806274
37 0.702620267868042
};
\addlegendentry{$\gpeak=6$}
\addplot [semithick, color0]
table {%
-37 1.02837598323822
-36 1.03338670730591
-35 1.04991090297699
-34 1.04456567764282
-33 1.03709673881531
-32 1.03621935844421
-31 1.02757930755615
-30 1.02994549274445
-29 1.02930998802185
-28 1.02663087844849
-27 1.01867437362671
-26 1.01767683029175
-25 1.02199053764343
-24 1.0162627696991
-23 1.01708841323853
-22 1.01848649978638
-21 1.01265954971313
-20 1.009108543396
-19 1.00846385955811
-18 1.00793218612671
-17 1.0066351890564
-16 1.00639390945435
-15 1.0064662694931
-14 1.00424194335938
-13 0.999924898147583
-12 1.00217866897583
-11 0.997429609298706
-10 0.997873783111572
-9 1.00075089931488
-8 0.993787050247192
-7 0.997596383094788
-6 0.998067617416382
-5 0.994593620300293
-4 0.993288636207581
-3 0.992273330688477
-2 0.99259352684021
-1 0.987436830997467
0 0.989230215549469
1 0.986879229545593
2 0.99111008644104
3 0.988133311271667
4 0.987615823745728
5 0.985229790210724
6 0.987941861152649
7 0.9871666431427
8 0.988391816616058
9 0.984438419342041
10 0.982549905776978
11 0.987545967102051
12 0.984489381313324
13 0.983137190341949
14 0.984145760536194
15 0.988579988479614
16 0.985767602920532
17 0.983724594116211
18 0.985374331474304
19 0.984719455242157
20 0.984309554100037
21 0.986033618450165
22 0.978580474853516
23 0.977412581443787
24 0.97972297668457
25 0.979795098304749
26 0.980057835578918
27 0.975641489028931
28 0.979750394821167
29 0.980135083198547
30 0.98473858833313
31 0.989875257015228
32 0.99312686920166
33 0.995208323001862
34 1.00295197963715
35 1.0019907951355
36 1.00795793533325
37 0.984476447105408
};
\addlegendentry{$\gpeak=9$}
\end{axis}

\end{tikzpicture}
	  \captionof{figure}{Mean energy of the clusters centers per subcarrier.}
	  \label{fig:energy_clusters}
	\end{minipage}%
	\hspace{20pt}
	\begin{minipage}{.45\textwidth}
	  \centering
\begin{tikzpicture}

\definecolor{color0}{rgb}{0.172549019607843,0.627450980392157,0.172549019607843}

\pgfplotsset{
    width=.88\textwidth,
    height=0.8\textwidth
}
\begin{axis}[
legend cell align={left},
legend style={
  fill opacity=0.8,
  draw opacity=1,
  text opacity=1,
  at={(0.98,0.80)},
  anchor=north east,
  draw=white!80!black
},tick align=outside,
tick pos=left,
x grid style={white!69.0196078431373!black},
xlabel={Subcarrier index $n$},
xmajorgrids,
xtick = {-30, -15, 0, 15, 30},
xmin=-40.7, xmax=40.7,
xtick style={color=black},
y grid style={white!69.0196078431373!black},
ylabel={$\EE_k \left[ \bar{\mathcal{V}}(n,k) \right]$},
ymajorgrids,
ymin=-0.00300030952203088, ymax=0.105235780536896,
ytick style={color=black}
]
\addplot [semithick, color0, dash pattern=on 1pt off 1pt]
table {%
-37 0.00236992677673697
-36 0.00399293610826135
-35 0.00527686718851328
-34 0.00654227565973997
-33 0.00835634022951126
-32 0.0100266812369227
-31 0.0121355373412371
-30 0.0150105534121394
-29 0.0180342942476273
-28 0.0210529379546642
-27 0.0238719061017036
-26 0.0282119456678629
-25 0.0323494896292686
-24 0.0349351167678833
-23 0.039652943611145
-22 0.0429932437837124
-21 0.0474329814314842
-20 0.0522184669971466
-19 0.0579137578606606
-18 0.0603289678692818
-17 0.0640129670500755
-16 0.0683849453926086
-15 0.0695051103830338
-14 0.0744373798370361
-13 0.0785176604986191
-12 0.0817599892616272
-11 0.0824424624443054
-10 0.0871515572071075
-9 0.0899104475975037
-8 0.089974507689476
-7 0.0924372300505638
-6 0.0945228263735771
-5 0.096869058907032
-4 0.0957691073417664
-3 0.0977316051721573
-2 0.0995995104312897
-1 0.10031595826149
0 0.09930120408535
1 0.0997341573238373
2 0.09739089012146
3 0.0977607667446136
4 0.0998202115297318
5 0.0939553007483482
6 0.096929520368576
7 0.0946988314390182
8 0.092244416475296
9 0.093160055577755
10 0.0874670594930649
11 0.0857717469334602
12 0.0831417292356491
13 0.0787613540887833
14 0.0768825337290764
15 0.0723550617694855
16 0.0695578157901764
17 0.0683194100856781
18 0.0628066807985306
19 0.058399360626936
20 0.0538759157061577
21 0.0495542883872986
22 0.0441190339624882
23 0.040836364030838
24 0.0370115116238594
25 0.0329110771417618
26 0.0280208420008421
27 0.0250614378601313
28 0.0212688408792019
29 0.0175448711961508
30 0.0152736501768231
31 0.0129267740994692
32 0.0104460101574659
33 0.00837888102978468
34 0.00703837536275387
35 0.00542258005589247
36 0.00428517814725637
37 0.00313353724777699
};
\addlegendentry{$\gpeak=4$}
\addplot [semithick, color0, dashed]
table {%
-37 0.00540816318243742
-36 0.00760700600221753
-35 0.0087707219645381
-34 0.0101925265043974
-33 0.011961636133492
-32 0.0136224031448364
-31 0.0145704858005047
-30 0.0156695917248726
-29 0.0185642577707767
-28 0.0195951163768768
-27 0.0208053849637508
-26 0.0214329399168491
-25 0.0228074565529823
-24 0.023183761164546
-23 0.024000121280551
-22 0.0245742071419954
-21 0.0255801770836115
-20 0.0251171216368675
-19 0.0259223114699125
-18 0.0250634998083115
-17 0.0250591095536947
-16 0.0251071471720934
-15 0.023878026753664
-14 0.0246168952435255
-13 0.0248988457024097
-12 0.0244314968585968
-11 0.0250933691859245
-10 0.0245676040649414
-9 0.0253265034407377
-8 0.0242699570953846
-7 0.0254705101251602
-6 0.0257171727716923
-5 0.0262017101049423
-4 0.0257603712379932
-3 0.0276981983333826
-2 0.02618701569736
-1 0.0272855181246996
0 0.0270138289779425
1 0.0278956852853298
2 0.0267101936042309
3 0.026178490370512
4 0.0269775446504354
5 0.0263022761791945
6 0.0266084391623735
7 0.0260528400540352
8 0.0250609815120697
9 0.0250132326036692
10 0.024156641215086
11 0.0243000909686089
12 0.0236383080482483
13 0.0244725719094276
14 0.0247382372617722
15 0.0236441232264042
16 0.0248508173972368
17 0.0235223341733217
18 0.0244376100599766
19 0.0252332091331482
20 0.0255083926022053
21 0.0245855841785669
22 0.0236639082431793
23 0.0239622481167316
24 0.0227193832397461
25 0.0219429358839989
26 0.0203659702092409
27 0.0195059422403574
28 0.0184330381453037
29 0.0170971639454365
30 0.0152290519326925
31 0.0140707874670625
32 0.0123738739639521
33 0.0114187002182007
34 0.00990207865834236
35 0.00808265805244446
36 0.0069289542734623
37 0.00444358214735985
};
\addlegendentry{$\gpeak=6$}
\addplot [semithick, color0]
table {%
-37 0.00194135005585849
-36 0.00241701910272241
-35 0.00305531197227538
-34 0.0032062167301774
-33 0.00337721267715096
-32 0.00345108122564852
-31 0.00371955242007971
-30 0.00357673550024629
-29 0.00376886129379272
-28 0.00361257651820779
-27 0.00358180934563279
-26 0.00347060197964311
-25 0.00345570337958634
-24 0.00340102636255324
-23 0.0036219151224941
-22 0.00348233059048653
-21 0.00350812147371471
-20 0.00342566007748246
-19 0.0033929068595171
-18 0.00355413882061839
-17 0.00342899421229959
-16 0.00342630315572023
-15 0.00352016324177384
-14 0.00343902269378304
-13 0.00363869313150644
-12 0.00349832838401198
-11 0.00345001299865544
-10 0.00332666980102658
-9 0.00373976957052946
-8 0.00342874601483345
-7 0.00358276604674757
-6 0.00354064558632672
-5 0.003543337341398
-4 0.00368084362708032
-3 0.00362831493839622
-2 0.00363948056474328
-1 0.00347705837339163
0 0.00351503281854093
1 0.00382779911160469
2 0.00363931455649436
3 0.00374700827524066
4 0.00359053118154407
5 0.00365619827061892
6 0.00366006209515035
7 0.00372651405632496
8 0.00369421951472759
9 0.00334116909652948
10 0.00366501370444894
11 0.00358625152148306
12 0.00352695817127824
13 0.00332213588990271
14 0.00339021370746195
15 0.00329783326014876
16 0.00334052392281592
17 0.00336855719797313
18 0.00351426145061851
19 0.00331861292943358
20 0.00337796402163804
21 0.00350360479205847
22 0.00343355815857649
23 0.00343952886760235
24 0.00339724565856159
25 0.00338888401165605
26 0.00337113277055323
27 0.00329481577500701
28 0.00342671573162079
29 0.00343230436556041
30 0.00333064468577504
31 0.00337376981042325
32 0.0032363859936595
33 0.00335650099441409
34 0.00291319796815515
35 0.00304419291205704
36 0.00233611464500427
37 0.00191951275337487
};
\addlegendentry{$\gpeak=9$}
\end{axis}

\end{tikzpicture}
	  \vspace{-20pt}
	  \captionof{figure}{Mean variance of the clusters per subcarrier.}
	  \label{fig:var_clusters}
	\end{minipage}
\end{figure}

To better understand the neural transmitter behavior, we index each cluster by a tuple $(n,k)$, where $n$ is the subcarrier index and $k\in\{0, \cdots, 2^K\}$ the index of the cluster for the $n^{\text{th}}$ subcarrier.
Let us denote by $\left\{ \bv^{(k)} \right\}_{1 \leq k \leq 2^K}$ the set of all possible vectors of bits indexed by their decimal representation, i.e., $\bv^{(0)} = [0, 0]\tp, \bv^{(1)} = [0, 1]\tp, \bv^{(2)} = [1, 0]\tp$, and $\bv^{(3)} = [1, 1]\tp$ for $K=2$.
We verified that each cluster corresponds to a unique vector of bits $\bv^{(k)}$, and the center of these clusters are represented by purple dots in Fig.~\ref{fig:clusters}.
For each E2E system, we define a new high-dimensional constellation $\bar{\mathcal{C}} \in \CC^{N \times 2^K}$ that comprises the centers of the $2^K$ clusters on all $N$ subcarriers.
We denote by $\bar{\mathcal{C}}(n, k) = \EE_{\xv_m} \left[ x_{m,n}^{(k)}\right]$ the $k^{\text{th}}$ constellation point in the $n^{\text{th}}$ subcarrier, where $x_{m,n}^{(k)}$ denotes the output of the neural transmitter for the RE $(m,n)$ when $\bv^{(k)}$ was given as input.
Similarly, we define $\bar{\mathcal{V}}\in \CC^{N \times 2^K}$, such that $\bar{\mathcal{V}}(n, k) = \EE_{\xv_m} \left[ x_{m,n}^{(k)} {x_{m,n}^{(k)}}^* \right]$ represents the variance of the cluster $(n,k)$.

Fig.~\ref{fig:energy_clusters} shows the mean energy of the constellation on each subcarrier for E2E systems trained with $\gpeak\in\{4, 6, 9\}~\si{dB}$ and a lax \gls{ACLR} constraint, corresponding to the three rows of Fig.~\ref{fig:clusters}.
We can verify that when the \gls{PAPR} constraint is lax ($\gpeak=9\si{dB}$), the energy is evenly distributed across the subcarriers.
On the contrary, the border subcarriers are given less energy when a harsh constraint is applied ($\gpeak=4\si{dB}$).
One explanation could be that the center subcarriers have longer wavelength, and therefore contribute less than their counterparts with shorter wavelengths. 
By focusing the available energy on these subcarriers, the neural transmitter jointly minimizes the probability of peaks in the analog waveform and decreases the number of \glspl{FBS} that have a significant impact on them.
Note that since the carrier frequency is typically much higher than the bandwidth of each subcarrier, the average power of the passband and baseband signals differs by a factor two but their maximum are approximately equal, which amounts to a roughly 3~\si{dB} difference between the passband and baseband \gls{PAPR}~\cite{eras2019analysis}.
Finally, the mean variance of the clusters on each subcarrier are plotted in Fig.~\ref{fig:var_clusters} for the same three systems.
One can observe that the clusters exhibit almost no variance with $\gpeak=9\si{dB}$, but a high variance in the center frequencies with $\gpeak=4\si{dB}$.
These high variances observed with harsh \gls{PAPR} constraints tend to indicate that the neural transmitter is able to slightly relocate the relevant \glspl{FBS} in order to minimize the waveform \gls{PAPR}.
Overall, the transmitter seems to focus its energy on central subcarriers to reduce the probability of peaks and the number of relevant \glspl{FBS}, and to adjust the positions of these \glspl{FBS}  to minimize the peak amplitudes.

\begin{figure}
	\centering
	\begin{minipage}{.45\textwidth}
	  \centering
\begin{tikzpicture}

\definecolor{color0}{rgb}{0.172549019607843,0.627450980392157,0.172549019607843}
\definecolor{color1}{rgb}{0.580392156862745,0.403921568627451,0.741176470588235}

\pgfplotsset{
    width=.88\textwidth,
    height=0.8\textwidth
}

\begin{axis}[
legend cell align={left},
legend style={fill opacity=0.8, draw opacity=1, text opacity=1, draw=white!80!black},
log basis y={10},
tick align=outside,
tick pos=left,
x grid style={white!69.0196078431373!black},
xlabel={ $e~[\si{\dB}]$},
xmajorgrids,
xmin=2, xmax=10,
xtick style={color=black},
y grid style={white!69.0196078431373!black},
ylabel={$\text{CCDF}_{|\alpha(t)|^2}(e)$ },
ymajorgrids,
ymin=5e-06, ymax=3,
ymode=log,
ytick style={color=black}
]
\addplot [semithick, color0, dash pattern=on 1pt off 1pt, mark=+, mark size=3, mark options={solid}]
table {%
0.0306621547788382 0.463842057142857
0.302457183599472 0.433853219047619
0.574252188205719 0.402547923809524
0.846047222614288 0.369996952380952
1.11784219741821 0.336168723809524
1.38963723182678 0.301184419047619
1.66143226623535 0.265173371428571
1.93322730064392 0.2283024
2.20502233505249 0.190936761904762
2.47681736946106 0.153442057142857
2.74861240386963 0.116525676190476
3.0204074382782 0.0815008380952381
3.29220247268677 0.0505213714285714
3.56399750709534 0.026621219047619
3.83579230308533 0.0114221333333333
4.1075873374939 0.00381154285714286
4.37938261032104 0.000903961904761905
4.65117740631104 0.000129142857142857
4.92297267913818 9.37142857142857e-06
5.19476747512817 3.04761904761905e-07
5.46656274795532 3.80952380952381e-08
};
\addlegendentry{E2E, $\gpeak=4~\si{dB}$}
\addplot [semithick, color1, dash pattern=on 1pt off 1pt, mark=o, mark size=2, mark options={solid}]
table {%
0.215409502387047 0.334035619047619
0.510280966758728 0.309861676190476
0.805152416229248 0.286007352380952
1.10002386569977 0.262591771428571
1.39489543437958 0.239740076190476
1.6897668838501 0.217626895238095
1.98463833332062 0.196333714285714
2.27950978279114 0.175950133333333
2.57438135147095 0.156632342857143
2.86925268173218 0.138449904761905
3.16412425041199 0.121488457142857
3.45899558067322 0.105796457142857
3.75386714935303 0.0914179809523809
4.04873847961426 0.078414780952381
4.34361028671265 0.0666984761904762
4.63848161697388 0.0562793523809524
4.93335294723511 0.0471129523809524
5.22822427749634 0.0391968761904762
5.52309608459473 0.0323596952380952
5.81796741485596 0.026524
6.11283874511719 0.0216392
6.40771055221558 0.0175647619047619
6.70258188247681 0.014241219047619
6.99745321273804 0.0114997714285714
7.29232454299927 0.00928251428571429
7.58719635009766 0.0074928380952381
7.88206768035889 0.00604148571428571
8.17693901062012 0.0048736380952381
8.47181034088135 0.00392731428571429
8.76668167114258 0.00315333333333333
9.06155395507812 0.00252057142857143
9.35642528533936 0.00199660952380952
9.65129661560059 0.00156765714285714
9.94616794586182 0.00121428571428571
10.241039276123 0.000925942857142857
10.5359106063843 0.000698133333333333
10.8307828903198 0.000514628571428571
11.1256542205811 0.000368152380952381
11.4205255508423 0.000258247619047619
11.7153968811035 0.000175619047619048
12.0102682113647 0.000114780952380952
12.305139541626 7.22285714285714e-05
12.6000108718872 4.31238095238095e-05
12.8948831558228 2.46857142857143e-05
13.189754486084 1.2952380952381e-05
13.4846258163452 6.8952380952381e-06
13.7794971466064 3.65714285714286e-06
14.0743684768677 1.75238095238095e-06
14.3692398071289 9.52380952380952e-07
14.6641111373901 3.80952380952381e-07
14.9589834213257 2.66666666666667e-07
15.2538547515869 1.52380952380952e-07
15.5487260818481 3.80952380952381e-08
};
\addlegendentry{Extracted const.}
\end{axis}

\end{tikzpicture}
	  \captionof{figure}{CCDF of the power of the E2E system trained for $\gpeak=4\si{dB}$ \mbox{($\bleak=-20\si{dB}$)} and of a conventional system using its extracted constellation.} 
	  \label{fig:ccdf_const}
	\end{minipage}%
	\hspace{20pt}
	\begin{minipage}{.45\textwidth}
	  \centering
\begin{tikzpicture}

\definecolor{color0}{rgb}{0.172549019607843,0.627450980392157,0.172549019607843}
\definecolor{color1}{rgb}{0.12156862745098,0.466666666666667,0.705882352941177}

\pgfplotsset{
    width=.88\textwidth,
    height=0.8\textwidth
}

\begin{axis}[
legend cell align={left},
legend style={fill opacity=0.8, draw opacity=1, text opacity=1, draw=white!80!black},
log basis y={10},
tick align=outside,
tick pos=left,
x grid style={white!69.0196078431373!black},
xlabel={$e~[\si{\dB}]$},
xmajorgrids,
xmin=2, xmax=10,
xtick style={color=black},
y grid style={white!69.0196078431373!black},
ylabel={$\text{CCDF}_{|\alpha(t)|^2}(e) $},
ymajorgrids,
ymin=5e-06, ymax=3,
ymode=log,
ytick style={color=black}
]
\addplot [semithick, color0, mark=+, mark size=3, mark options={solid}]
table {%
0.101464577019215 0.393805180952381
0.396041452884674 0.367264685714286
0.690618336200714 0.340525904761905
0.985195159912109 0.313687314285714
1.27977204322815 0.286951695238095
1.57434892654419 0.260429942857143
1.86892580986023 0.234163238095238
2.16350269317627 0.208384266666667
2.45807957649231 0.183314361904762
2.75265645980835 0.159071733333333
3.04723334312439 0.1358256
3.34181022644043 0.113744342857143
3.63638710975647 0.0931153142857143
3.93096399307251 0.0740155047619048
4.22554063796997 0.0566458666666667
4.52011775970459 0.0411877333333333
4.81469440460205 0.0278787428571429
5.10927152633667 0.0169222476190476
5.40384817123413 0.00841066666666667
5.69842529296875 0.00289321904761905
5.99300193786621 0.000699428571428571
6.28757905960083 0.00012247619047619
6.58215570449829 1.37904761904762e-05
6.87673282623291 1.06666666666667e-06
};
\addlegendentry{E2E, $\gpeak=6~\si{dB}$}
\addplot [semithick, color1, mark=square, mark size=2, mark options={solid}]
table {%
0.324601471424103 0.34156
0.698699653148651 0.310030095238095
1.07279777526855 0.278978285714286
1.4468959569931 0.248454476190476
1.82099413871765 0.219090666666667
2.19509243965149 0.190974857142857
2.56919050216675 0.164407619047619
2.94328880310059 0.139785523809524
3.31738686561584 0.116993904761905
3.6914849281311 0.0963424761904762
4.06558322906494 0.0781241904761905
4.43968152999878 0.0619611428571429
4.81377935409546 0.0480864761904762
5.1878776550293 0.0365234285714286
5.56197595596313 0.0269889523809524
5.93607425689697 0.0193977142857143
6.31017208099365 0.0135622857142857
6.68427038192749 0.00915580952380952
7.05836868286133 0.00598476190476191
7.43246698379517 0.00373371428571429
7.80656480789185 0.00226742857142857
8.18066310882568 0.00131047619047619
8.55476093292236 0.000723047619047619
8.92885971069336 0.000386666666666667
9.30295753479004 0.000187809523809524
9.67705535888672 8.87619047619048e-05
10.0511541366577 3.69523809523809e-05
10.4252519607544 1.1047619047619e-05
10.7993507385254 4.19047619047619e-06
11.1734485626221 1.9047619047619e-06
11.5475463867188 1.14285714285714e-06
11.9216451644897 7.61904761904762e-07
};
\addlegendentry{Baseline, $R=0$}
\end{axis}

\end{tikzpicture}
	  \captionof{figure}{CCDF of the power of the E2E system trained with $\gpeak=6\si{dB}$ \mbox{($\bleak=-20\si{dB}$)} and of the baseline using 16-QAM modulation and no \gls{PRT}.}
	  \label{fig:ccdf}
	\end{minipage}
\end{figure}

In order to verify this claim, the  \gls{CCDF}  of the ratio between the instantaneous and average power of the transmitted signal is defined as
\begin{align}
	\text{CCDF}_{|\alpha(t)|^2}(e) = P \left( |\alpha(t)|^2 > e \right) \approx P \left( \left| \frac{\underline{z}_{m,t}}{\EE\LSB \underline{z}_{m,t} \RSB } \right| ^2 > e \right)
\end{align}
and is approximated by sending $10000$ batches of $1000$ \glspl{RG}, each \gls{RG} $i$ being composed of $N=75$ subcarriers and $M=14$ OFDM symbols $\underline{z}_{m,t}^{[i]} $ oversampled with a factor $5$.
The \gls{CCDF} of two systems are presented in Fig.~\ref{fig:ccdf_const}, the first one corresponds to the first row of Fig.~\ref{fig:clusters}, i.e., it is trained with the harshest \gls{PAPR} constraint ($\gpeak=9\si{dB}$) and a lax \gls{ACLR} constraint ($\bleak=-20\si{dB}$).
The second one is a conventional system that uses the constellation $\bar{\mathcal{C}}$ extracted from the aforementioned E2E system (and represented by purple dots in the first row of Fig.~\ref{fig:clusters}).
The goal of this comparison is to evaluate the effect of the adjustment of the \glspl{FBS}  positions from the clusters centers operated by the transmitter.
It can be seen that the \gls{CCDF} of the E2E system is drastically lower than the one of the conventional system using the extracted constellation, indicating that the \gls{PAPR} reduction is indeed performed through the \glspl{FBS}  position adjustments.

Finally, the \glspl{CCDF} of the E2E system trained with $\gpeak=6\si{dB}$ (and a lax \gls{ACLR} constraint) and of a conventional \gls{QAM} system are compared in Fig.~\ref{fig:ccdf}.
The E2E system corresponds to the second row of Fig.~\ref{fig:clusters}, and the two compared systems are respectively highlighted by a black circle and a black square in Fig.~\ref{fig:rates}.
On can see that the \gls{PAPR} minimization process operated by the neural transmitter is particularly effective, as the \gls{CCDF} of the E2E system is significantly lower despite the two systems enabling similar rates.
Finally, it is also interesting to note that the \gls{CCDF} of the system using the extracted constellation in Fig.~\ref{fig:ccdf_const} is higher than the one of the 16-QAM system in Fig.~\ref{fig:ccdf}.
This indicates that the underlying learned constellation alone has worse \gls{PAPR} characteristics than a standard \gls{QAM}, demonstrating the efficiency of the positional adjustments enabled by the neural transmitter.

\section{Conclusion} 
\label{sec:conclusion}

In this paper, we proposed an end-to-end learning approach to design OFDM waveforms that meet specific constrains on the envelope and spectral characteristics.
To that aim, the transmitter and receiver are modeled as two \glspl{CNN} that perform high-dimensional modulation and demodulation, respectively
The training procedure first requires all optimization constraints to be expressed as differentiable functions that can be minimized through \gls{SGD}.
Then, a constrained optimization problem is formulated and solved using the augmented Lagrangian method.
We evaluated the proposed approach on the learning of OFDM waveforms that maximize an information rate of the transmission while satisfying \gls{PAPR} and \gls{ACLR} constraints.
Simulations were performed using \gls{3GPP}-compliant channel models, and results show that the optimization procedure is able to design waveforms that satisfy the \gls{PAPR} and \gls{ACLR} constraints.
Moreover, the end-to-end system enables up to 30\% higher throughput than a close to optimal implementation of a \gls{TR} baseline with similar \gls{ACLR} and \gls{PAPR}.
Evaluation insights revealed that the neural transmitter achieves \gls{PAPR} and \gls{ACLR} reduction through a subcarrier-dependent filtering, an uneven energy distribution across subcarrier, and a positional readjustment of each constellation point.
Future research directions include the training and evaluation of end-to-end systems with other optimization constraints and with larger \glspl{RG} on channel model corresponding to medium- and high-mobility scenarios, possibly with hardware imperfections.

\bibliographystyle{IEEEtran}
\bibliography{IEEEabrv, bib_abrv, bibliography}

\end{document}